# Understanding the attitudes, knowledge sharing behaviors and task performance of core developers: A longitudinal study

Sherlock A. Licorish and Stephen G. MacDonell
*Department of Information Science*
*University of Otago*
*PO Box 56, Dunedin 9054*
*New Zealand*
sherlock.licorish@otago.ac.nz, stephen.macdonell@otago.ac.nz

## Abstract

**Context:** Prior research has established that a few individuals generally dominate project communication and source code changes during software development. Moreover, this pattern has been found to exist irrespective of task assignments at project initiation. **Objective**: While this phenomenon has been noted, prior research has not sought to understand these dominant individuals. Previous work considering the effect of team structures on team performance has found that core communicators are the gatekeepers of their teams' knowledge, and the performance of these members was correlated with their teams' success. Building on this work, we have employed a longitudinal approach to study the way core developers' attitudes, knowledge sharing behaviors and task performance change over the course of their project, based on the analysis of repository data. **Method**: We first used Social Network Analysis (SNA) and standard statistical analysis techniques to identify and select artifacts from ten different software development teams. These procedures were also used to select central practitioners among these teams. We then applied psycholinguistic analysis and directed content analysis (CA) techniques to interpret the content of these practitioners' messages. Finally, we inspected these core developers' activities as recorded in system change logs at various points in time during systems' development. **Results**: Among our findings, we observe that core developers' attitudes and knowledge sharing behaviors were linked to their involvement in actual software development and the demands of their wider project teams. However, core developers appeared to naturally possess high levels of insightful characteristics, which became evident very early during teamwork. **Conclusion:** Project performance would likely benefit from strategies aimed at surrounding core developers with other competent communicators. Core developers should also be supported by a wider team who are willing to ask questions and challenge their ideas. Finally, the availability of adequate communication channels would help with maintaining positive team climate, and this is likely to mitigate the negative effects of distance during distributed developments.

**KEYWORDS:** Software Development, Core Developers, Psycholinguistics, Content Analysis, Attitudes, Knowledge Sharing, Task Performance

## 1. INTRODUCTION

Previous research has established that a few individuals in a team generally dominate project communication and source code changes during software development [1-4]. Evidence has also shown that, even in environments with fixed task assignments, specific individuals circumvent these pre-set arrangements to occupy the center of their teams' activities [5]. Such patterns have been studied previously in other disciplines [6, 7], and early works investigating the effect of this phenomenon have shown that the existence of these centralized patterns involving core group members is a positive sign for team performance [8]. In similarly seminal work, Leavitt established that central individuals are vital to their teams' performance as they coordinate information flow. Central individuals are also *seen as* project leaders, whether or not they are the formal or nominal leaders [9], and groups with central coordinators experience higher levels of group organization and task performance (in terms of speed when completing tasks) [8].

While there is therefore strong interest in identifying patterns within software teams' communication and coordination practices, there has been comparatively little effort directed toward understanding why these patterns exist or how they emerge. Questions related to how core members share knowledge over their project, the initial arrangements that lead to core members becoming hubs in their teams, and how the attitudes and traits these practitioners exhibit might be linked to their involvement in task changes, have not been answered. Such explorations could provide insights into the peculiarities of software team dynamics, inform appropriate team configurations, and enable the early identification of 'software gems' – exceptional practitioners in terms of both task and team performance.

In previous work we used psycholinguistics and content analysis techniques to study the roles and behaviors of core developers and uncovered that these practitioners worked across multiple roles, and were indeed crucial to their teams' organizational, intra-personal and inter-personal processes [10]. Additionally, we noted that although these individuals were highly task- and achievement-focused, they were also largely responsible for maintaining positive team atmosphere [10]. While we established that core developers exhibited significantly different attitudes and behaviors to their 'regular' counterparts during team work, this prior study took a static, single snapshot view of the software project teams considered. Additionally, our goal in that prior work was primarily to examine *if* core developers behaved differently to their less active counterparts.

In building on that work we now answer the questions noted above in this paper. In the current work we employ a multi-stage approach to study the ways in which core developers' attitudes and knowledge sharing behaviors change over time. In order to do so we examine actual software artifacts as against the more frequent use of surveys and interviews [11]. Our analysis was conducted in several steps, described in detail in Section 3. Briefly, in the first step repository data was mined and artifacts from multiple software development teams were explored using social network analysis (SNA) and standard statistical techniques. This enabled us to select our cases and to detect patterns around core developers, and so informed the design of the later steps involving deeper linguistic analysis techniques and directed content analysis (CA).

Our work makes several contributions. In terms of methodological contribution, we demonstrate the value of employing contextual analysis techniques to understand internal software team processes. We extend software engineering (SE) theory and explain how and why core developers become knowledge hubs, we provide understandings for the way core developers' attitudes and knowledge sharing behaviors are linked to their involvement in software tasks, and we outline several avenues for future research. We also provide recommendations for software project governance and show how the outcomes of our work have implications for team strategies. Finally, we provide suggestions for the enhancement of collaboration and process support tools.

In the next section (Section 2) we present our theoretical background and survey related works; this leads to our specific research questions. We then provide details of our research setting and measures in Section 3, introducing our techniques and procedures in this section. In Section 4 we present our results and we analyze and discuss our findings. Section 5 outlines the implications of our results, and in Section 6 we consider our study's limitations. Finally, in Section 7 we draw our conclusions.

## 2. THE STUDY OF TEAM COMMUNICATION

Previous work has established that the intricacies of team dynamics can be revealed by studying members' communication [12]. Research has also uncovered linkages between informal hierarchical communication structures and team performance for geographically distributed teams [13]. Furthermore, team communication has been linked to coordination efficiency [14] and the quality of software output [15]. Thus, studying the details in team communication can provide valuable insights into the human processes involved during software development, including the reasons for, and consequences of, communication and coordination actions.

Given this, software repositories and software history data have emerged as valuable sources of interaction and communication evidence [16]. Research findings drawn from works examining such sources are particularly valid if the data represents the primary means of interaction and so captures team processes during software development [10]. Accordingly, previous researchers have exploited process artifacts such as electronic messages, change request histories, bug logs and blogs to provide unique perspectives on the activities occurring during the software development process [3, 17]. In particular, open source software (OSS) repositories and archives recording software developers' textual communication activities have increasingly provided researchers with opportunities to study software practitioners' behaviors [18, 19].

For instance, Abreu and Premraj [20] analyzed the Eclipse mailing list and found that increases in communication intensity coincided with higher numbers of bug-introducing changes, and that developers communicated most frequently at release and integration time. Bird et al. [4] employed clustering algorithms to study CVS records and mailing lists and concluded that the more software development an individual does the more coordination and controlling activities they must undertake. These observations are supported by Cataldo et al. [3], whose SNA study found that central individuals contributed the most during software development. The Debian mailing list was used by Sowe et al. [21] to observe knowledge sharing among developers. These authors found that no specific individual dominated knowledge sharing activities in the Debian project.

Works such as those of Bacchelli et al. [22] and Antoniol et al. [23] have used rather more complex techniques to analyze email and bug description information. In linking email communications to changes in source code using regular expressions and other information retrieval approaches, Bacchelli et al. [22] found that the analysis approach using regular expressions in emails outperformed more complex probabilistic and vector space models. Through the use of decision trees, naïve Bayes classifiers and logistic regression, Antoniol et al. [23] were also able to classify bugs based on specific terms used in the textual descriptions of such tasks.

In reviewing these two streams of work, it is evident that some researchers have looked to infer the semantics of practitioners' dialogues from the text they communicated (e.g., [22, 23]), while others have provided deductions based on communication frequency information [4, 20]. While text analysis methods and their associated tools have been used previously to understand and predict some aspects of software development [24, 25], only a few studies in this domain have considered examining software teams' internal behavioral processes as represented in their members' textual communications. This is in spite of the fact, as noted by Bacchelli et al. [22], that natural language analysis techniques have proved to be effective in generating understandings of software developers' attitudes when applied to their language processes.

For instance, in analyzing the communication of the developers involved in the Apache project, Rigby and Hassan [12] uncovered that once the two most active developers signaled their intentions to leave the project their communications became more negative and instructive, they spoke mostly in the future tense, and they communicated with less positive emotions, when compared to their earlier communications. In our own work examining three different IBM Rational Jazz project teams we also found slight variances in behaviors among those undertaking different forms of software task [25]. As noted in Section 1, as a first step to understanding the true role of active communicators, we used linguistic analysis and content analysis techniques to study these individuals' messages and change histories [10]. We found that core communicators were also core developers (coinciding with the findings of Cataldo and Herbsleb [1]); and that these practitioners were extremely valuable to their teams in terms of maintaining team harmony, knowledge sharing, team organization and task performance. Such findings support the utility of natural language analysis techniques for understanding human processes.

Questions over reliability and validity have also been raised in relation to studies analyzing OSS repositories in terms of arriving at generalizable conclusions regarding software process issues. Research evidence has reported poor data quality in some repositories of OSS projects [4, 26]. For instance, in their study of the Apache mailing list Bird et al. [4] found it difficult to uniquely identify developers' records due to the volume of email addresses and aliases these individuals used. Further confounding issues may also be encountered when studying OSS projects because anyone is able to post messages and report bugs to their associated mailing lists, whether or not those individuals are contributing to the project [27] or even have a full understanding of the project. Given these issues, coupled with the potential value of studying team interactions, and the gaps identified above, it is imperative that we examine the contextual interactions and engagements of contemporary software practitioners, using representative systems, if we are to comprehend the unique nature of these teams [28]. We therefore extend the work we conducted previously [10] in this study, and look in more detail at core developers' attitudes and knowledge sharing behaviors and their involvement in software development tasks using a longitudinal approach. We outline our specific research questions under the following two subsections (subsection 2.1 and subsection 2.2).

## 2.1. Attitudes and Knowledge Sharing Behaviors

Central communicators have been previously referred to as active communicators, core communicators, core members and core developers [2, 12, 29, 30], in light of findings that clearly demonstrate that individuals' engagements in communication and development are related. In this study we use the term "core developers" to refer to those that are both actively involved in communication and task performance. This classification has been used previously by those studying both OSS and commercial software teams [2, 30].

It was previously noted that core developers contribute significantly to maintaining team harmony, knowledge sharing, team organization and task performance [10]. However, it is not clear how these individuals contribute to their teams' processes over the course of their project, and how their organizational, inter-personal and intra-personal competencies sustain their project's health. Additionally, there is uncertainty around what team conditions, and over which project phase(s), core developers are most important to their teams. Previous work has shown that practitioners' interaction patterns differ over the duration of their project [1, 31, 32], and so longitudinal studies should help us to understand software team dynamics more fully. Evidence of how practitioners interact over the course of their project could inform targeted team strategies and phase-specific interventions. In fact, previous calls for such investigations of team dynamics have been made [13], as the static view does not reveal fully what actually happens over the duration of a software development project.

Linguistic studies have shown that while an individual's language use is stable over time, the way individuals communicate is also influenced by their context and local settings [33]. Thus, an individual may express happiness and satisfaction in their communication if they are fulfilled, while the opposite may be observed if they are dissatisfied [34]. In software development settings, negative and cynical team behaviors can have a negative impact on team harmony and cohesion [35, 36]. This will in turn affect team performance [37]. The opposite is likely to occur in more optimistic environments where teams share a single vision. This is a particularly critical issue for distributed software development contexts, where individuals are already affected by distance and have few if any opportunities to engage in face-to-face communication [38, 39] which has been shown to stimulate trust [40-42]. Given that core developers occupy the center of coordination action, are seen as the project's leaders (whether or not they are the assigned leaders [13]), and that they coordinate information flow and knowledge sharing [9], an understanding of core developers' attitudes and knowledge sharing behaviors will be useful in informing strategies aimed at maintaining an optimistic and positive team climate, and ultimately, positive team performance. On this basis we ask our first two research questions:

*RQ1. Do core developers' attitudes change as their project progresses?*

*RQ2. What knowledge sharing behaviors do core developers exhibit over the course of their project?*

## 2.2. The Relationship Between Attitudes and Knowledge Sharing Behaviors and Task Performance

Research examining software teams' communication has focused primarily on the use of social network related measures (e.g., centrality and closeness [4, 5, 13, 30, 32, 43]). In fact, the studies that have concluded that just a few individuals contribute the most to communication and task changes have generally used frequency-based techniques [1-3], and while there have been some efforts to understand the characteristics of core developers [1], these works did not probe the reasons underlying the existence of this phenomenon. While such techniques do enable the detection of certain patterns, and so provide a partial understanding of software teams' behavioral processes, there are limitations on the effectiveness of these approaches in informing our understanding of the deeper psychosocial nature of team dynamics [10, 28].

Although we now know that core developers are invaluable to their teams [10] beyond just liaison and task change roles [1], open questions remain around when these individuals are more or less likely to perform, and during which phases these practitioners are most influential – details that are likely to provide valuable insights for software project governance. In the same vein, revealing the process of how developers become part of the core group should help project leaders to identify and encourage software gems very early in their project. Some developers may occupy natural roles, such that, regardless of the project environment, these individuals may function in a certain way based on their natural preferences [44]. On the other hand, others may evolve into specific roles given their teams' demands and their specific task assignments [45]. Knowledge and awareness of these different developers and the way they work should help project leaders to identify software development leaders early, and in assembling high performing and cohesive teams. We first study the way core developers contribute to task performance over the course of their projects by answering the following question, prior to examining the way these members' attitudes and knowledge sharing behaviors are linked to their task performance:

*RQ3. How do core developers contribute to task performance over their project?*

Knowledge sharing studies have shown that the willingness of individuals and teams to actively participate in knowledge sharing and contribute to team performance is linked to multiple factors [46]. For instance, knowledge sharing has been linked to social motivation (e.g., trust [47, 48]), rewards and incentives [49], cognitive factors [50], and other organizational reasons [51]. While there is some uncertainty around the effects of incentives and rewards on individuals' active participation in knowledge sharing [52], social motivation theory has proved to be generally effective for predicting participation in knowledge sharing [47, 48, 53]. According to social motivation theory, teams' inter-personal interactions and norms have an impact on individual members' motivation to perform [53]. Thus, certain supportive behavioral norms at the team level are likely to encourage an individual's performance [54]. This position may be especially valid given that knowledge sharing is a social process [52].

Such theories motivate our interest in understanding how core developers' attitudes and contributions of knowledge are linked to their performance (and that of their teams), and when their teams are most likely to benefit from their knowledge and experiences. Such insights would potentially be useful for understanding the specific traits of less prudent team members that are likely to complement these core individuals. Additionally, such understandings could inform specific project arrangements that can enhance the satisfaction of core developers. Furthermore, these revelations could inform decisions regarding how software teams should be staffed during core developers' less productive periods, thereby supporting overall project management and maintaining team performance. These potential outcomes inform our last two research questions:

*RQ4. Are core developers' contributions to task performance linked to their attitudes?*

*RQ5. Are core developers' contributions to task performance linked to their contributions of knowledge?*

## 3. RESEARCH SETTING

We examined development artifacts from a specific release (1.0.1) of Jazz (based on the IBM$^{R}$ Rational$^{R}$ Team Concert$^{TM}$ (RTC) [1]), a fully functional environment for developing software and for managing the entire software development process [55]. The software includes features for work planning and traceability, software builds, code analysis, bug tracking and version control in one system [56]. Changes to source code in the Jazz environment are permitted only as a consequence of a work item (WI) being created beforehand, such as a defect, a task or an enhancement request. Defects are actions related to bug fixing, whereas design documents, documentation or support for the RTC online community are labeled as tasks (although we refer to them here as 'support tasks' in order to differentiate with our general use of 'task' in the paper). Enhancements relate to the provision of new functionality or the extension of system features (and a history log is maintained for each WI). Team member communication and interaction around WIs are captured by Jazz's comment or message functionality. During development at IBM, project communication, the content explored in this study, was enforced through the use of Jazz itself [57].

The release of the Jazz environment to which we were given access comprised a large amount of process data collected from distributed software development and management activities across the USA, Canada and Europe. In Jazz each team has multiple individual roles, with a project leader responsible for the management and coordination of the activities undertaken by the team (and team members may also work across project teams – see Figure 1 for illustration) [30]. All Jazz teams use the Eclipse-way methodology for guiding the software development process [55]. This methodology outlines iteration cycles that are six to eight weeks in duration, comprising planning, development and stabilizing phases, generally conforming to agile principles. Builds are executed after project iterations. All information for the software process is stored in a server repository, which is accessible through a web-

[1] IBM, the IBM logo, ibm.com, and Rational are trademarks or registered trademarks of International Business Machines Corporation in the United States, other countries, or both

based or Eclipse-based (RTC) client interface [58]. This consolidated data storage and enforced project control mean that the data in Jazz is much more complete and representative of the software process than that in many OSS repositories. We provide details of our data extraction process and metrics definitions in the following two subsections (subsection 3.1 and subsection 3.2).

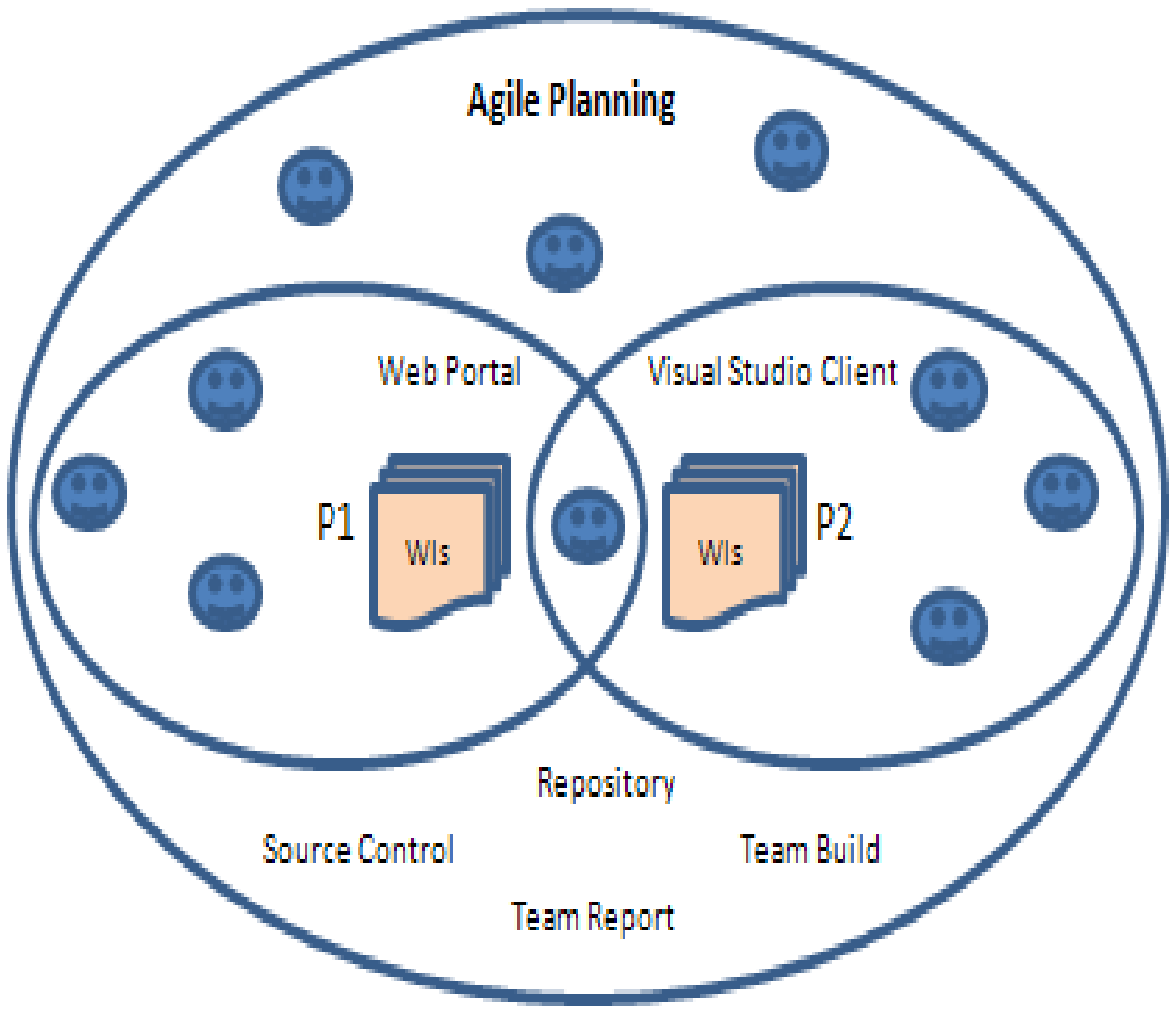

Figure 1. Project teams' arrangement in Jazz

### 3.1. Data Extraction

Although an investigation of data mining is beyond the scope of this paper we briefly report here the relevant steps performed in this project in terms of extracting, preparing and exploring the data under observation [59]. Data cleaning, integration and transformation techniques were utilized to maximize the representativeness of the data under consideration and to help with the assurance of data quality, while exploratory data analysis (EDA) techniques were employed to investigate data properties and to facilitate anomaly detection [60]. Through these latter activities we were able to identify all records with inconsistent formats and data types, for example: an integer column with an empty cell. We wrote scripts to search for these inconsistent records and tagged those for deletion. This exercise allowed us to identify and delete 122 records (out of 36,672) that were of inconsistent format. We also wrote scripts that removed all HTML tags and foreign characters (as these would have confounded our analysis).

We leveraged the IBM Rational Jazz Client API to extract team information and development and communication artifacts from the Jazz repository. These included (in addition to the WIs discussed above):

- Project Workspace/Area – each Jazz team is assigned a workspace. The workspace (or project/team area P*n*) contains all the artifacts belonging to the specific team (see Figure 1 for a conceptual illustration).
- Contributors and Teams – a contributor is a practitioner contributing to one or more software features; multiple contributors in a Project Area form a team.
- Comments or Messages – communication around WIs is captured by Jazz comment functionality. Messages ranged from as short as one word (e.g., "thanks"), up to 1055 words representing multiple pages of communication.

We extracted the relevant information from the repository and selected all the artifacts belonging to ten different project teams (out of 94) for analysis. As noted earlier, these project teams were employed across locations in North America and Europe; however, we did not consider the specific team location as the unit of analysis in this work. We are aware that cultural differences and distance (geographical and temporal) may directly affect software development teams' performance [39], and these conditions may also have an impact on team members' behaviors - which in turn may lead to performance issues [61]. However, previous research examining the effects of cultural differences in global software teams has found few cultural gaps and behavioral differences among software practitioners from, and operating in, Western cultures (the setting for the teams studied in this work), with the largest negative effects observed between Asian and Western cultures [39]. Thus, while we acknowledge that different team dynamics may be evident in collocated settings, our focus in this work was not on the entire team, or the effect of coordination in collocated versus distributed work (refer to Espinosa et al. [62] for further details on this subject), but on understanding core software developers' attitudes, knowledge sharing behaviors and task performance. Accordingly, the specific location of developers was not considered as a major influence on the pattern of results in this work.

The artifacts selected formed a purposive rather than random sample. Table 1 shows that the selected project areas (comprising project teams' artifacts) represented both information-rich and information-rare cases in terms of WIs and messages. Project areas had tasks covering as few as two iterations to as many as 17 iterations (refer to Table 1), with varying levels of communication density. Density varies between 0 and 1 [63], where a task or individual that attracted interaction from all the members in a team would have a density of 1, while those with no interaction would have a density of 0 (e.g., in terms of team members' density measures, a practitioner that communicated on 20 out of their team's 50 tasks would have a density of 20/50 = 0.4).

The selected project artifacts amounted to 1201 software development tasks comprising 692 defects, 295 support tasks and 214 enhancements, carried out by a total of 394 contributors working across the ten project teams (and comprising 146 distinct members working across five roles), with 5563 messages exchanged around the 1201 tasks. Although each of the ten teams had a slightly different balance in terms of role membership (refer to column four in Table 1), as the data were analyzed, it became clear that the cases selected were representative of those in the repository in terms of team members' engagement, as we reached data saturation [64] after analyzing the third project case. Our use of SNA to initially explore the project's communication from task-based social networks [58] showed a similar graph to that in Figure 2 for all of the ten project teams (note the dense communication segments – those with high incidence of communication – for developers 12065 and 13664 respectively). Figure 2 demonstrates a typical Jazz team's task-based directed social network, where network edges belonging to distinct contributors on individual software tasks are merged and color coded; edge color moved from red to brown (between one to five messages), brown to green (between six to ten messages) and then to a more pronounced green (ten or more messages). The network vertices represented either a class image denoting a task or a contributor's unique identification number. In addition to the network visualizations, all ten project teams had similar profiles for network density (between 0.02 and 0.14) and closeness (between 0 and 0.06) – these were validated through formal statistical testing [65]. This consistency in SNA measures suggests that the teams selected were indeed relatively homogenous (refer to Licorish [65] for further details around our application of SNA to study software teams' collaboration in general).

## 3.2. Description of Measures

We present our measures and our units of analysis in this section. First, we outline the techniques that were used to identify and select the teams' core developers (subsection 3.2.1). Second, we describe the procedure that was used for separating team artifacts into the different time phases (subsection 3.2.2). We then define and describe the measures that were used for analyzing attitudes in the third subsection (subsection 3.2.3). Our approach to measuring knowledge sharing behaviors is explained in the fourth subsection (subsection 3.2.4). Finally, we describe approaches that have been utilized previously for measuring task performance and explain the specific approach used in this study (subsection 3.2.5).We used core developers as our units of analysis and made comparisons across project phases.

### *3.2.1. Selecting Core Developers*

We first created a baseline level of developer contribution using an approach similar to that used by Crowston et al. [29], and selected all practitioners whose communication density measure was $\geq 0.33$ (i.e., they communicated on a third or more of their teams' project tasks), and labeled them as *core developers* (see our earlier work [10] for further details). Fourteen contributors across the ten project teams met this selection criterion – a smaller number than preferred in terms of supporting our intended analysis. We therefore relaxed our initial requirements and selected the top two communicators of each team, following the approach used by Rigby and Hassan [12], which increased the total number of top members by six (to comprise 20 practitioners in total, including 15 distinct core developers). Eight out of the 15 distinct core group members were programmers, along with five team leaders and two project managers [10]. As a second step to validating that these members were indeed core developers, analysis confirmed that these individuals had initiated more than 41% of all software tasks, made more than 69% of the changes to these tasks and resolved nearly 75% of all software tasks undertaken by their teams (refer to [10] for further details). On the basis that the selected practitioners had the highest levels of involvement in team communication and task changes we believe that these members can indeed be considered to be core developers – those most engaged and active in their teams' performance [10]. However, the potential for such figures to be confounded by a project's organizational and managerial structures has encouraged us to conduct the more in depth analyses using qualitative methods reported here (refer to Section 4.2 for further details).

### *3.2.2. Separating Project Phases*

Software tasks were planned and executed over multiple iterations for each project area (shown as P1 to P10 in Table 1). However, the number of iterations varied across the project areas (e.g., P3's tasks were completed in two iterations, whereas P5's tasks were executed over 17 iterations – refer to the sixth column of Table 1). In order to normalize these artifacts, we therefore time-sliced each team's tasks and artifacts into four equal quarters (start, early-mid, late-mid, and end) to enable us to study changes in core practitioners' attitudes, knowledge sharing behaviors and task performance over the duration of their project [25]. So for example, P1 artifacts were divided into four blocks of 76 days each, while P3 artifacts were divided into three 15-day blocks and one 14-day block (refer to Table 1 for details). Differences in the number and type of tasks performed (described above) may be thought to potentially affect the patterns of teams' communication that are observed. We in fact discovered that IBM Rational Jazz teams communicated more messages when solving enhancement features (5.2 messages on average) than they did around support tasks (4.8 messages on average) and defects (4.4 messages on average). Similarly, Jazz teams communicated the most at project start and project completion. These patterns were consistent across the teams in Table 1, regardless of the teams' focus or priorities and the number of participants and roles involved [66]. However, core developers (in all of the teams in Table 1) were equally active on all software tasks (also discussed in subsection 3.2.1 above) [10]. This evidence suggests that differences in the number and type of tasks performed may not have a major effect on the patterns of attitudes and knowledge sharing behaviors that these core developers exhibit over the course of their project. Furthermore, in alignment with this study's focus, our deeper analyses enable us to explore and understand these potential differences further.

Table 1. Summary statistics for the selected Jazz project teams

| Team ID | Task (WI) Count | Software Tasks (Project/ Team Area) | Total Contributors – Roles | Total Messages | Period (days) – Iterations |
|---|---|---|---|---|---|
| P1 | 54 | User Experience – tasks related to UI development | 33 – 18 programmers, 11 team leads, 2 project managers, 1 admin, 1 multiple roles | 460 | 304 - 04 |
| P2 | 112 | User Experience – tasks related to UI development | 47 – 24 programmers, 14 team leads, 2 project managers, 1 admin, 6 multiple roles | 975 | 630 - 11 |
| P3 | 30 | Documentation – tasks related to Web portal documentation | 29 – 12 programmers, 10 team leads, 4 project managers, 1 admin, 2 multiple roles | 158 | 59 - 02 |
| P4 | 214 | Code (Functionality) – tasks related to development of application middleware | 39 – 20 programmers, 11 team leads, 2 project managers, 2 admins, 4 multiple roles | 883 | 539 - 06 |
| P5 | 122 | Code (Functionality) – tasks related to development of application middleware | 48 – 23 programmers, 14 team leads, 4 project managers, 1 admin, 6 multiple roles | 539 | 1014 - 17 |
| P6 | 111 | Code (Functionality) – tasks related to development of application middleware | 25 – 11 programmers, 9 team leads, 2 project managers, 3 multiple roles | 553 | 224 - 13 |
| P7 | 91 | Code (Functionality) – tasks related to development of application middleware | 16 – 6 programmers, 7 team leads, 1 project manager, 1 admin, 1 multiple roles | 489 | 360 - 11 |
| P8 | 210 | Project Management – tasks under the project managers' control | 90 – 29 programmers, 24 team leads, 6 project managers, 2 admins, 29 multiple roles | 612 | 660 - 16 |
| P9 | 50 | Code (Functionality) – tasks related to development of application middleware | 19 – 10 programmers, 3 team leads, 4 project managers, 2 multiple roles | 254 | 390 - 10 |
| P10 | 207 | Code (Functionality) – tasks related to development of application middleware | 48 – 22 programmers, 12 team leads, 2 project managers, 1 admin, 11 multiple roles | 640 | 520 - 11 |
| ∑ | **1201** | | **394 contributors,** comprising 175 programmers, 115 team leads, 29 project managers, 10 admins, 65 multiple roles | **5563** | |

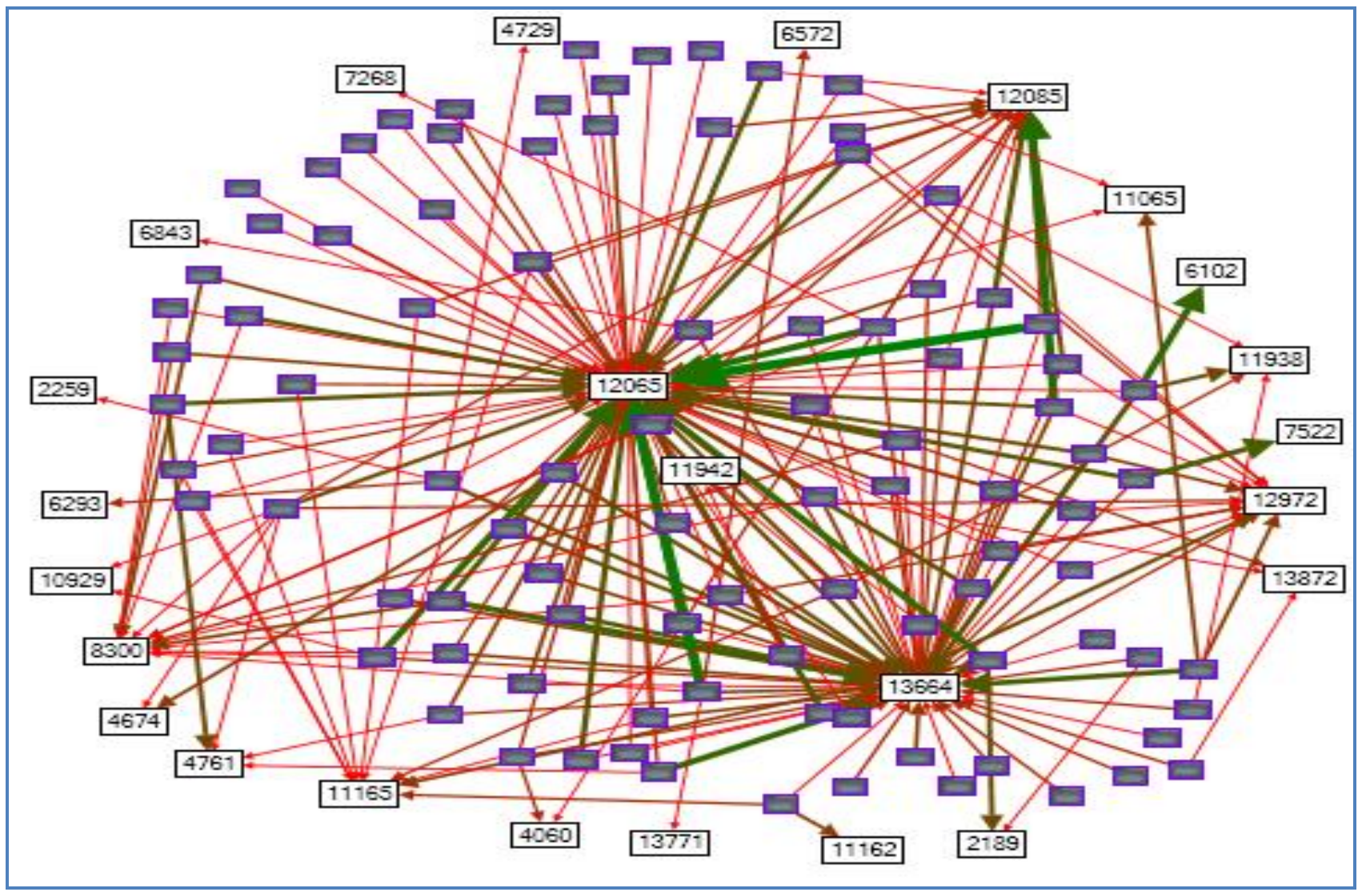

Figure 2. Directed network graph for a sample Jazz team showing highly dense network segments for practitioners "12065" and "13664"

### 3.2.3. *Measuring Attitudes*

Language use has been studied extensively across a range of social contexts [67-74]. These works have all provided evidence in support of the viewpoint that there are unique variations in individuals' linguistic styles from situation to situation, and linguistic analysis of textual communication can reveal much about those who are communicating. Researchers considering software developers' language processes have also conducted multiple forms of textual analysis to understand team processes, in support of this viewpoint [2, 12, 22, 75-77]. That said, the actual approaches that are used to study software practitioners' linguistic processes from their textual communication vary from study to study (e.g., [12, 22, 75]).

In following the lead of previous work [12, 78], we employed the Linguistic Inquiry and Word Count (LIWC) software tool in our analysis of core developers' attitudes. The LIWC is a software tool created after four decades of research using data collected across the USA, Canada and New Zealand [73]. This tool captures over 86% of the words used during conversations (around 4500 words). Written text is submitted as input to the tool in a file that is then processed and summarized based on the LIWC tool's dictionary. Each word in the file is searched for in the dictionary, and specific type counts are incremented based on the associated word category (if found), after which a percentage value is calculated by aggregating the number of words in each linguistic category over all words in the messages. For example, if there were 10 instances of words belonging to the "I" dimension (refer to Table 2) in a message with a length of 200 words, then the percentage value for the "I" dimension would be (10/200=)5.0%. The different categories (or dimensions) in the LIWC output summary are said to capture the attitudes of individuals by assessing the words they use [73, 78]. We provide a summary of the LIWC linguistic categories that were considered in our analysis in Table 2. The selection of these categories was informed by theories in social psychology [79], psycholinguistics [80] and role theories [81] (refer to the fourth column of Table 2 for details).

To illustrate how attitudes are measured using the LIWC tool's output, consider the following sample comment:

> "*We* are aiming to have all the patches ready by the end of this release; this will provide *us* some space for the next one. Also, *we* are extremely confident that similar bug-issues will not appear in the future."

In this comment the author is expressing optimism that the team will succeed, and in the process finish ahead of time and with acceptable quality standards. In this quotation, the words "we" and "us" (shown in italics) are indicators of team or collective focus, "all", "extremely" and "confident" are associated with certainty, while the words "some" and "appear" are indicators of tentative processes. As per the approach outlined above, the LIWC tool's output for the "we" dimension or collective focus for this comment (refer to Table 2) would be (3/39=)7.7%. Words in the comment above such as "bug-issues" and "patches" are not included in the LIWC dictionary, and so would not affect the context of its output – whether it was to indicate a fault in software code or a problem with one's immunity to, and treatment for, a disease. Although these omissions may be thought to represent a confounding factor in our analysis, we already know that the circumstance is software development and that this context is consistent. Additionally, what *is* of interest, and *is* being captured by the tool, is evidence of attitudes in this context. Furthermore, as noted above, previous work has provided confirmation of the utility of the LIWC tool for examining attitudes [12, 25], and we have also triangulated our linguistic findings in this study through the application of complementary analysis methods.

### 3.2.4. *Measuring Knowledge Sharing Behaviors*

We studied the knowledge sharing behaviors of core developers through their messages using a directed CA approach, employing a hybrid classification scheme adapted from related prior work [84, 85]. Use of a directed CA approach is appropriate when there is scope to extend or complement existing theories around a phenomenon [86], and so suited our explorations of the Jazz core developers' knowledge sharing behaviors. The directed content analyst approaches the data analysis process using existing theories to identify key concepts and definitions as initial coding categories. In our case, we used theories examining knowledge sharing behaviors expressed during textual interaction [84, 85] to inform our initial categories (resulting in scales 1-8 in Table 3). Henri and Kaye [84] and Zhu [85] used Bretz' [87] three-stage theory of interactivity and the group interaction theory of Hatano and Inagaki [88] and Graesser and Person [89] respectively to study teams' interaction. Henri and Kaye [84]'s coding instrument was created to observe five areas of interactivity: participation, social, interaction, cognitive and meta-cognitive communication, while Zhu [85]'s social interaction protocol looked to classify vertical or horizontal interaction. Vertical interaction is characterized by communication where group members seek answers or solutions to problems from capable members, while horizontal interaction involves the strong assertion of ideas, answers, information, discussions, comments, reflections and scaffolding.

Should existing theories prove insufficient to capture all relevant insights during preliminary CA coding, new categories and subcategories should be created [86]. Both authors of this work and two other experienced qualitative coders first classified a random sample of 5% of the core developers' communications in a preliminary coding phase to verify the suitability of the initial protocol that was created (comprising scales 1-8 in Table 3). During this exercise we found that some aspects of Jazz core developers' interactions were not captured by the first version of our protocol (e.g., Instructions and Gratitude were missing). The Henri and Kaye [84] and Zhu [85] protocols did not previously cover these dimensions of interaction. During this pilot coding exercise we also found that practitioners in Jazz communicated multiple ideas in their messages. Thus, we segmented the communication using the sentence (or *utterance*) as the unit of analysis. We extended the protocol resulting in new scales 9 to 13 in Table 3 (note: these scales emerged in the order in which they appear in the Table), after which the first author and the two experienced coders recoded 709 messages. As many codes as were necessary were assigned to utterances that demonstrated multiple forms of interaction – or knowledge sharing behaviors – and all coding differences were discussed and resolved by consensus (see Section 4.2 for details). We achieved 81% inter-rater agreement between the three coders as measured using Holsti's coefficient of reliability measurement (C.R) [90]. This represents excellent agreement between coders and suggests that a consistent and reliable approach was being taken.

Table 2. Summary of linguistic categories

| **Linguistic Category** | **Abbreviation (Abbrev.)** | **Examples** | **Reason for Inclusion** |
|---|---|---|---|
| Pronouns | I | I, me, mine, my, I'll, I've, myself, I'm | Elevated use of first person plural pronouns (we) is evident during shared situations, whereas, relatively high use of self-references (I) has been linked to individualistic attitudes [82]. Use of the second person pronoun (you) may signal the degree to which members rely on (or delegate to) other team members [80]. |
| | we | we, us, our, we've, lets, we'd, we're, we'll | |
| | you | you, your, you'll, you've, y'all, you'd, yours, you're | |
| Cognitive language | insight | think, consider, determined, idea | Software teams were previously found to be most successful when many group members were highly cognitive and were natural solution providers [83]. These traits are also linked to effective task analysis and brainstorming capabilities. |
| | discrep | should, prefer, needed, regardless | |
| | tentat | maybe, perhaps, chance, hopeful | |
| | certain | definitely, always, extremely, certain | |
| Work and Achievement related language | work | feedback, goal, boss, overtime, program, delegate, duty, meeting | Individuals most concerned with task completion and achievement are said to reflect these traits during their communication. Such individuals are most concerned with task success, contributing and initiating ideas and knowledge towards task completion [81]. |
| | achieve | accomplish, attain, closure, resolve, obtain, finalize, fulfill, overcome, solve | |
| Leisure, social and positive language | leisure | club, movie, entertain, gym, party, jog, film | Individuals that are personal and social in nature are said to communicate positive emotion and social words and this trait is said to contribute towards an optimistic group climate [81]. Leisure related language may also be an indicator of a team-friendly atmosphere. |
| | social | give, buddy, love, explain, friend | |
| | posemo | beautiful, relax, perfect, glad, proud | |
| Negative language | negemo | afraid, bitch, hate, suck, dislike, shock, sorry, stupid, terrified | Negative emotion may affect team cohesiveness and group climate. This form of language shows discontent and resentment [79]. |

#### *3.2.5. Measuring Task Performance*

Various approaches have been used over many years to measure individual-level performance in software tasks. Productivity-related measures such as lines of code per unit of effort [91], time taken to complete development tasks [37] and the number of task changes completed [1] are among those used previously to measure performance. Along with others, Cataldo and Herbsleb [1] argued that measures based on lines of code may not be reliable in instances where there is variability in developers' coding styles (i.e., some developers are more verbose than others). The time taken to complete development tasks may vary for developers when there are many feature inter-dependencies (e.g., a developer may start working on a feature that needs to use classes that are under development by another developer, and thus may be delayed). We therefore use the number of task changes as indicative of the performance of these core developers in software tasks [1]. A developer was considered to change a task if they created, modified, or resolved that task (as has been utilized in prior studies [1, 30]). We previously examined these measures separately in order to scrutinize them closely (refer to Licorish and MacDonell [10]) to mitigate for the potential that these core developers may have possibly made minor changes or worked on potential duplicate bugs that would falsely indicate that they were productive, and in the process, affect the reliability of our task performance metric [10]. Our results, as far as these quantitative measures were concerned, confirmed that core developers were extremely active in creating tasks, modifying these tasks and resolving these tasks, and particularly when these measures were compared to those for their less active counterparts [10]. However, as noted in Section 3.2.1, such figures may also be confounded by a project's organizational and managerial structures; thus, our in-depth analysis of core developers' messages is used here (as was evaluated previously [10]) to triangulate these outcomes. The results and analysis from these and other enquiries are provided next.

## 4. RESULTS AND ANALYSIS

As noted in Section 2, we have previously established that core developers expressed significantly different attitudes to their less active counterparts, and these individuals were also responsible for solving most of their teams' software tasks [10]. These findings were revealed by studying a single snapshot of the core developers' project artifacts over the entire development cycle (as represented in the release).

Here we conduct temporal analyses of these practitioners' attitudes, knowledge sharing behaviors and task performance to answer the research questions outlined above (in Section 2).

Table 4 provides a summary of the numbers and percentages of messages contributed by the core developers over the four phases of each of the ten teams in the Jazz project. In total, 2565 messages were contributed by the core developers (of the total 5563 messages noted in Table 1). These practitioners typically communicated most in the early and middle phases of the project (see the measures for P1, P2, P5, P7 and P8 in Table 4). Previously we found that, overall, IBM Rational Jazz project teams communicated most in the first and last phases while addressing their development tasks [25, 66], and that the less active developers communicated more towards project completion. This finding was also noted by Cataldo and Herbsleb [1], who uncovered that technical dependencies resulted in increased levels of communication for some of the less active developers at various times of the project.

In the following subsection (subsection 4.1) we provide the results of our linguistic analyses of core developers' messages in order to answer RQ1. We then provide the results of our directed CA and related analysis to address RQ2 in subsection 4.2. Results and analysis regarding core developers' task changes are provided in subsection 4.3, and these enable us to answer RQ3. We subsequently examine the relationships between the linguistic analyses and task change results in subsection 4.4, and in the process we answer RQ4. Finally, the relationships between the directed CA and task change results along with appropriate discussions are presented in subsection 4.5 in order to answer RQ5.

## 4.1. Attitudes (RQ1)

In order to answer RQ1 we aggregated core developers' messages for the ten project teams (for each phase noted in Table 4), and analyzed these messages according to the 13 linguistic dimensions shown in Table 2. First, measures for the 13 individual linguistic dimensions were checked for normality [92] over the four project phases (start, early-mid, late-mid, and end) for the ten teams using the Shapiro-Wilk test. We found homogeneity of variance over the different project phases using Levene's test for equality of error variances for all of the linguistic dimensions. Ten of the thirteen linguistic dimensions (i.e., I, you, insight, discrep, tentat, certain, work, achieve, social and negemo – refer to Table 2) were found to be normally distributed across the four project phases. Closer examination of the standardized skewness coefficient (i.e., the skewness value divided by its standard error) and standardized kurtosis coefficient (i.e., the kurtosis value divided by its standard error) for these ten linguistic dimensions further established that these distributions conformed to the normality assumption. Measures for collective, leisure and positive emotion language (refer to Table 2) failed the Shapiro-Wilk normality test. We therefore conducted two-way ANOVA tests to check for differences in mean linguistic scores over project phases for the ten linguistic dimensions with normal distributions, and we conducted equivalent non-parametric Kruskal-Wallis tests for the other three non-normally distributed linguistic dimensions just mentioned [93].

Table 3. Coding categories for measuring knowledge sharing behavior

| Scale | Category | Characteristics and Example |
|---|---|---|
| 1 | Type I Question | Ask for information or requesting an answer – "Where should I start looking for the bug?" |
| 2 | Type II Question | Inquire, start a dialogue - "Shall we integrate the new feature into the current iteration, given the conflicts that were reported when we attempted same last week?" |
| 3 | Answer | Provide answer for information seeking questions - "The bug was noticed after integrating code change 305, you should start debugging here." |
| 4 | Information sharing | Share information – "Just for your information, we successfully integrated change 305 last evening." |
| 5 | Discussion | Elaborate, exchange, and express ideas or thoughts – "What was most intriguing about solving this bug is not how bugs may exist within code that went through rigorous testing... but how refactoring reveals bugs even though no functional changes are made." |
| 6 | Comment | Judgmental – "I disagree that refactoring may be considered the ultimate test of code quality." |
| 7 | Reflection | Evaluation, self-appraisal of experience – "I found solving the problems in change 305 to be exhausting, but I learnt a few techniques that should be useful in the future." |
| 8 | Scaffolding | Provide guidance and suggestions to others – "Let's document the procedures that were involved in solving this problem 305; it may be quite useful for new members joining the team in the future." |
| 9 | Instruction/ Command | Directive – "Solve task 234 in this iteration, there is quite a bit planned for the next." |
| 10 | Gratitude/ Praise | Thankful or offering commendation – "Thanks for your suggestions, your advice actually worked." |
| 11 | Off task | Communication not related to solving the task under consideration – "How was your weekend?" |
| 12 | Not Coded | Communication that does not fit codes 1 to 11, or 13. |
| 13 | Apology | Expressing regret or remorse – "I am very sorry for the oversight, and I am quite unhappy with the failure this has caused." |

Table 4. Numbers (and percentages) of messages communicated by core developers

| Team ID | Phase | | | | ∑ |
|---|---|---|---|---|---|
| | **start (%)** | **early-mid (%)** | **late-mid (%)** | **end (%)** | |
| P1 | 51 (19.8) | 96 (37.2) | 55 (21.3) | 56 (21.7) | **258** |
| P2 | 138 (26.8) | 184 (35.6) | 106 (20.5) | 89 (17.2) | **517** |
| P3 | 25 (52.1) | 11 (22.9) | 7 (14.6) | 5 (10.4) | **48** |
| P4 | 83 (28.2) | 73 (24.8) | 72 (24.5) | 66 (22.5) | **294** |
| P5 | 38 (20.5) | 28 (15.1) | 73 (39.5) | 46 (24.9) | **185** |
| P6 | 77 (19.8) | 96 (24.7) | 93 (23.9) | 123 (31.6) | **389** |
| P7 | 75 (23.7) | 82 (26.0) | 89 (28.2) | 70 (22.2) | **316** |
| P8 | 22 (16.3) | 28 (20.7) | 52 (38.5) | 33 (24.4) | **135** |
| P9 | 42 (36.2) | 33 (28.5) | 21 (18.1) | 20 (17.2) | **116** |
| P10 | 106 (34.5) | 72 (23.5) | 63 (20.5) | 66 (21.5) | **307** |
| ∑ | **657 (25.6)** | **703 (27.4)** | **631 (24.6)** | **574 (22.4)** | **2565** |

Table 5. Descriptive statistics for core developers' linguistic measures across the project phases

| Abbrev. | Mean (% use per phase) | | | | Median (% use per phase) | | | | Standard Deviation (std dev) | | | |
|---|---|---|---|---|---|---|---|---|---|---|---|---|
| | **start** | **early-mid** | **late-mid** | **end** | **start** | **early-mid** | **late-mid** | **end** | **start** | **early-mid** | **late-mid** | **end** |
| I | 6.8 | 7.6 | 7.5 | **7.9** | 6.6 | 9.1 | 8.9 | 7.2 | 4.4 | 4.7 | 4.4 | 3.5 |
| we | 2.9 | 2.8 | 2.4 | **4.1** | 3.0 | 2.3 | 2.1 | 2.9 | 1.7 | 2.7 | 1.3 | 3.4 |
| you | 2.9 | 3.4 | **3.5** | 2.4 | 2.8 | 3.5 | 3.0 | 2.6 | 1.0 | 1.4 | 1.7 | 1.9 |
| insight | 5.7 | **6.7** | 6.4 | 5.7 | 5.1 | 5.3 | 6.2 | 5.4 | 3.4 | 4.2 | 2.2 | 1.4 |
| discrep | 6.4 | **6.5** | 5.7 | 5.8 | 6.9 | 5.6 | 5.9 | 5.1 | 2.4 | 3.0 | 3.2 | 4.6 |
| tentat | **5.7** | 5.6 | 4.9 | 5.4 | 5.9 | 5.2 | 4.2 | 5.0 | 2.7 | 2.8 | 3.5 | 3.1 |
| certain | 2.3 | 1.9 | **2.6** | 2.4 | 2.8 | 1.9 | 2.5 | 2.4 | 1.3 | 0.8 | 1.5 | 1.7 |
| work | 11.8 | 11.0 | **14.0** | 13.5 | 10.5 | 10.2 | 14.4 | 11.4 | 3.1 | 3.9 | 2.1 | 7.3 |
| achieve | 10.5 | 10.2 | 10.5 | **10.7** | 10.6 | 9.7 | 10.8 | 9.4 | 3.7 | 4.5 | 3.3 | 5.3 |
| leisure | 3.1 | 2.6 | **3.3** | 2.7 | 3.1 | 2.1 | 2.6 | 1.7 | 1.2 | 2.1 | 1.7 | 2.8 |
| social | **14.8** | 12.9 | 12.7 | 12.7 | 14.9 | 12.3 | 13.3 | 13.2 | 3.1 | 3.5 | 4.3 | 4.2 |
| posemo | 15.7 | 17.9 | **18.4** | 17.0 | 9.9 | 11.8 | 13.6 | 12.3 | 15.2 | 14.8 | 12.8 | 15.2 |
| negemo | 3.1 | 2.7 | 3.6 | **4.8** | 2.8 | 2.9 | 3.4 | 3.9 | 1.8 | 1.5 | 1.6 | 4.9 |

Overall, although we were able to observe differences in the way core developers used the different language dimensions over time, these differences were not statistically significant ($p > 0.05$). For instance, Table 5 shows that core developers used lower levels of individualistic language at the start of their project (mean 6.8%, median 6.6%, std dev 4.4%), and use of this language form increased as the project progressed to completion (mean 7.9%, median 7.2%, std dev 3.5%). Table 5 also shows that core developers were less collective in the early phases of their project (mean 2.9%, median 3.0%, std dev 1.7%), and these practitioners tended to be most collective towards project completion (mean 4.1%, median 2.9%, std dev 3.4%). Use of work-related language exhibited a similar trend – core developers were highly work focused (frequently using words like "feedback", "goal" and "delegate") towards the end of the Jazz project (mean 13.5%, median 11.4%, std dev 7.3%). Additionally, core developers used a high amount of social language (e.g., give, buddy, love) throughout their project, but became less social as their project progressed (means: start 14.8%, early-mid 12.9%, late-mid 12.7%, and end 12.7%). Finally, Table 5 illustrates that while negative language use (e.g., "afraid", "hate", "dislike") was low overall for core developers (mean 3.5%, median 3.2%, std dev 2.6%), these practitioners expressed negative emotion mostly towards project completion (mean 4.8%, median 3.9%, std dev 4.9%). Refer to Table 5 for further descriptive statistics for core developers' linguistic measures across project phases. We now examine these results in relation to theory, and in the process answer RQ1.

*Discussion of RQ1. Do core developers' attitudes change as their project progresses?* **For the most part, the sample of Jazz core developers studied here communicated** **consistent attitudes over the duration of their project, but overall they appeared most cognitive at project initiation, and exhibited the most individualistic and negative attitudes towards project closure.** Overall, certain attitudes were more pronounced in specific project phases (refer to boldface mean values in Table 5). For instance, the Jazz core developers we studied became more self-focused as their project progressed, as shown in their increasing use of individualistic language (e.g., "I", "me", "my") over the course of their project. However, these individuals also exhibited higher levels of collective attitudes (using higher amounts of "we", "us", "our") towards project completion (noted for the measures for Abbrev. "we" in Table 5). Those exhibiting individualistic behaviors are said to affect team spirit and these traits have a negative effect on team cohesion [81], while the opposite is shown for those that are collective [82]. We hypothesize that the challenges of release pressure were possibly responsible for the higher levels of self-focused behavior at these times. Although these core developers communicated the least at project completion, and so the high levels of self-focus may have minimal negative impact on their team at this time, this form of attitude may be particularly undesirable during the early-mid and late-mid phases, when core developers are most engaged in the work of their team. Notwithstanding our small sample size (15 distinct practitioners) and the limitations of the psycholinguistic technique used in this work to study core developers' attitudes, we suggest that a project manager observing such trends should be vigilant and encourage more active contributions from the wider team during these times, given that core developers occupy the center of coordination action and their project teams' knowledge processes [9].

Core developers expressed the most cognitive attitudes (observed from the elevated measures for Abbrev. "insight" (insightful language), "discrep" (discrepancy language), "tentat" (tentative language) and "certain" (certainty language) in Table 5) at the start of their project, at a time when project features were being initiated. These findings are positive, given the need for project leaders to be perceptive and insightful (as seen for the use of words such as "think", "consider", "determine" and "idea" under the insight dimension in Table 2) during early project scoping activities. Members of the wider Jazz teams may benefit from these practitioners' higher levels of insight at this time, and so it may be prudent for project managers to implement strategies aimed at encouraging the engagement of less active developers at project inception. We observe that while the sample of Jazz core developers studied was highly task-focused [81] overall (refer to measures for Abbrev. "work" and "achieve" in Table 5), such organizational traits [10] were most evident in the latter project phases. These findings are particularly informative given core developers' lower levels of communication towards project completion (refer to measures in "end" column of Table 4). As noted previously, while project release pressure is held to promote such urgency among the core developers, and so it is understandable that these members would use higher amounts of "feedback", "attain", "resolve", "closure" and "solve" words at this time (refer to Table 2 for further examples under the "Work and Achievement" linguistic category), this finding supports the view that these practitioners were also motivated to see their project through even when they were not entirely 'in control' [35].

Relatively speaking, although we did not uncover statistically significant differences in the way core developers expressed attitudes over the course of their project, we observed in Table 5 that these members were also most negative (used higher amounts of "dislike", "suck", "stupid") towards project closure (see higher measures for Abbrev. "negemo" in the end phase). These findings coincide with our results for the expression of individualistic attitudes in Table 5. Both individualistic attitudes [82] and negative behavior are counterproductive for team work [79]. However, positive and social language use (e.g., "give", "love", "beautiful", "perfect") is an indicator of team friendliness [81], and Table 5 shows that core developers also used significant amounts of these forms of language throughout their project (see measures for Abbrev. "social" and "posemo"). The use of social language was especially pronounced in the early phases for core developers, a time when inter-personal skills are critical to team formation and establishing team dynamics [94]. These higher levels of social and positive processes may offset the more cynical attitudes. However, as noted above, keen participation (including clear communication) and availability of the wider team may go some way to reducing tension and coordination breakdowns [15] and enhance team spirit around project completion for core developers. These analyses are triangulated in the following section.

## 4.2. Knowledge Sharing (RQ2)

In order to answer RQ2 we conducted contextual analysis using directed CA to study core developers' knowledge sharing behaviors (refer to Table 3 for coding categories). As noted in Section 3.2.4, we coded all the messages that were contributed by the core developers of project teams P1, P7 and P8 (see Table 4 for details). We deliberately selected these three project areas as they represented different task portfolios, and we had previously found differences among project teams working on different types of tasks [25]. Thus, our contextual analysis here triangulates these prior results. From the 709 messages we recorded 2191 utterances/codes (P1 = 648 codes, P7 = 1245 codes, and P8 = 298 codes). We use percentages in Figures 3 and 4 to visualize the differences in core developers' knowledge sharing contributions to their teams' knowledge pools during the four different phases of their project. Figure 3 (a) provides an aggregated summary of the core developers' knowledge sharing interactions during the Jazz project. We previously found Information sharing, Discussion, Scaffolding, Comments and Instructions to dominate core developers' discourses [10], and the results that are revealed in this study confirm and extend these findings. In looking at the results from a temporal perspective, Figure 3 (b) shows that core developers contributed 60% of their knowledge during the middle phases of their project. A Chi-square test confirmed that there were significant differences in the level of contribution of core developers over the different project phases ($X^2 = 63.237$, df = 36, $p < 0.01$).

We consider in detail the nature of core developers' knowledge sharing interactions during their project by examining the categories of their interactions as outlined in Table 3 over the four project phases, start, early-mid, late-mid and end (refer to the graphs in Figure 4 (a-d)). We plot categories that are conceptually similar (e.g., Questions and Answers; Information sharing, Discussion and Scaffolding; and so on) in Figure 4 (a-d) so that the variances in the different scales are visible (this approach is also used in Figure 5) [84, 85]. This partitioning is slightly different to that in Table 3 and Figure 3 which reflects the actual ordering of the way codes were revealed during our analysis. Our directed CA results reveal that although core developers did not ask many questions overall (see Figure 3 (a)), Questions increased towards project completion: 34.6% and 30.6% of Type I and II Questions were asked in the last project phase (see Figure 4 (a)). Figure 4 (a) also shows that the majority of core developers' Answers to their teams' questions were provided in the middle phases (34.0% and 33.2% in early-mid and late-mid phases, respectively). Figure 4 (b) further shows that core developers shared most Information (33.8%), expressed more ideas (Discussion) (30.9%) and offered the most suggestions (Scaffolding) (31.4%) to their teams during the later middle phase of their project.

In fact, similar to our previous findings [10] we observe here that Information sharing dominated core developers' interactions (refer to Figure 3 (a)).The overall trend of the graphs in Figure 4 (b) indicates that core developers were most engaged in the middle (early-mid and late-mid) phases of their project. In looking at Comments in Figure 4 (c) we observe that there was a high contribution of this form of expression by core developers in the second project phase (more than 35%), but use of this language form remained relatively stable in the last two project phases (23.1% recorded in both phases). Figure 4 (c) shows that core developers were most Reflective towards project completion (32.2% of this language form was used in the end phase), and although core developers did not contribute a large amount of Off task communication or Gratitude (see Figure 3 (a)), these forms of utterances were also used most in the middle phases by these practitioners (see Figure 4 (d)). We did not plot the frequencies for messages Not Coded and Apology utterances in Figure 4 (as were shown in Figure 3 (a)), since only four and six codes (of the 2191 codes) were recorded to these categories, respectively.

We examined the knowledge sharing interaction measures for the core developers of the individual project teams (Figure 5) to see if the results above would hold across these members when they were undertaking their specific task types. For Type I and II Questions, Figure 5 (a) confirms that core developers asked more questions towards project completion when they were working in P1 (30.0% and 24.3% in the end phase) and P8 (66.7% and 45.5% in the end phase), while questions fluctuated, and were highest in the late-mid phase, for the core developers of P7 (42.9% and 30.0% in the late-mid project phase). In Figure 5 (a) the pattern for Answers is similar to the overall measures presented in Figure 4 (a), where core developers communicated most of this form of utterance in the middle phases (45.2% in the early-mid project phase for P1, 34.1% in the late-mid phase for P7 and 52.6% in the late-mid phase for P8). A similar trend is maintained for Information sharing, ideas (Discussion) and suggestions (Scaffolding) in Figure 5 (b): core developers tended to be most actively engaged using these types of utterances during the middle phases (early-mid and late-mid) of their project. Figure 5 (c) shows that the pattern of Comments remained similar to the overall pattern for teams P1 (61.5% contributed in the early-mid phase) and P7 (37.9% contributed in the early-mid phase), whereas this form of language use was most pronounced during the last project phase for core developers working on P8 (44.4% contributed during this period). Similarly, Figure 5 (c) shows that the core developers of P8 were most reflective at project completion (in alignment with the overall pattern of results); however, those of the other teams (P1 and P7) tended to be most reflective in the first and second project phases (50.0% contributed in the early-mid phase for P1 and 35.9% for the start phase of P7). Figure 5 (c) and Figure 5 (d) show that Instructions, Gratitude and Off task communications held the same pattern as that of the overall results across all three project teams for core developers. Overall, most of the results for core developers' knowledge sharing interaction in their individual teams held similar patterns to those that were evident when these members' interactions were aggregated (presented above). We now analyse these results in relation to theory to answer RQ2.

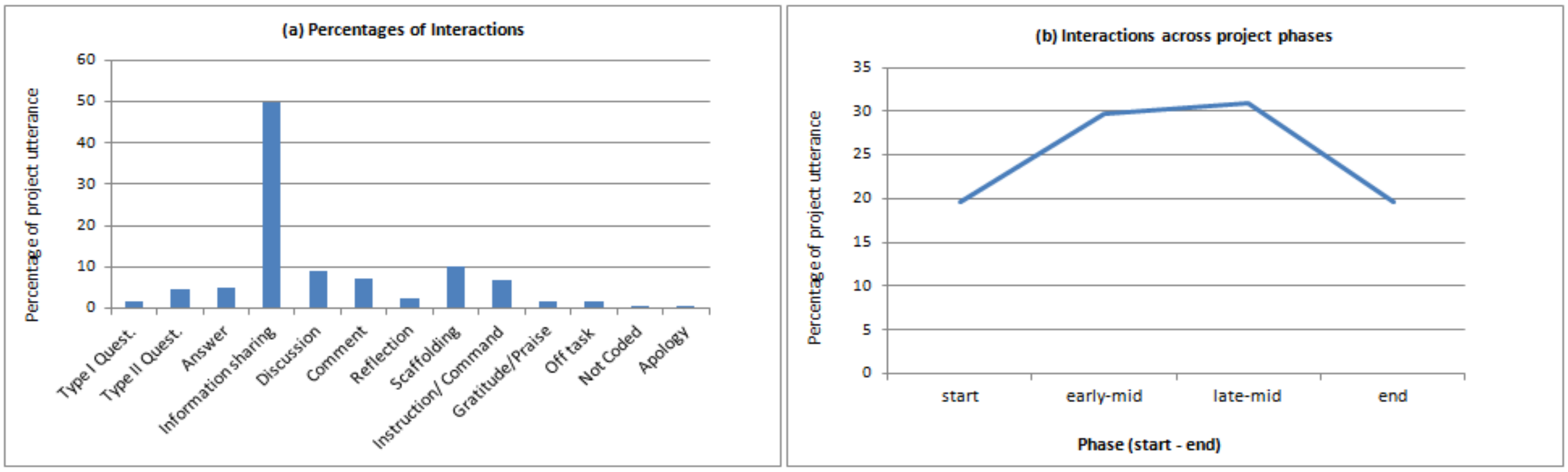

Figure 3. Aggregated interactions for core developers

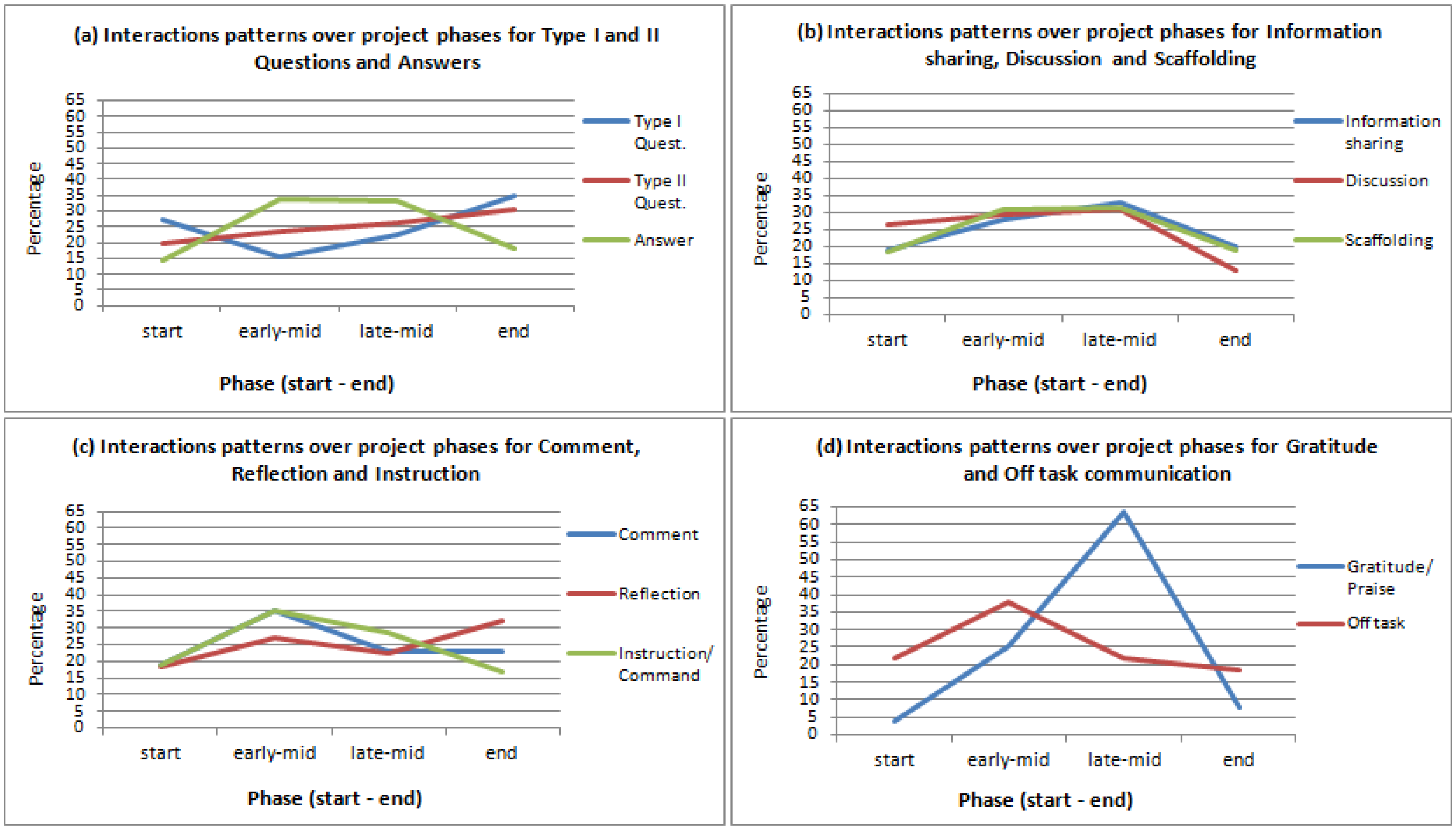

Figure 4. Detailed interactions of core developers over project phases (aggregated)

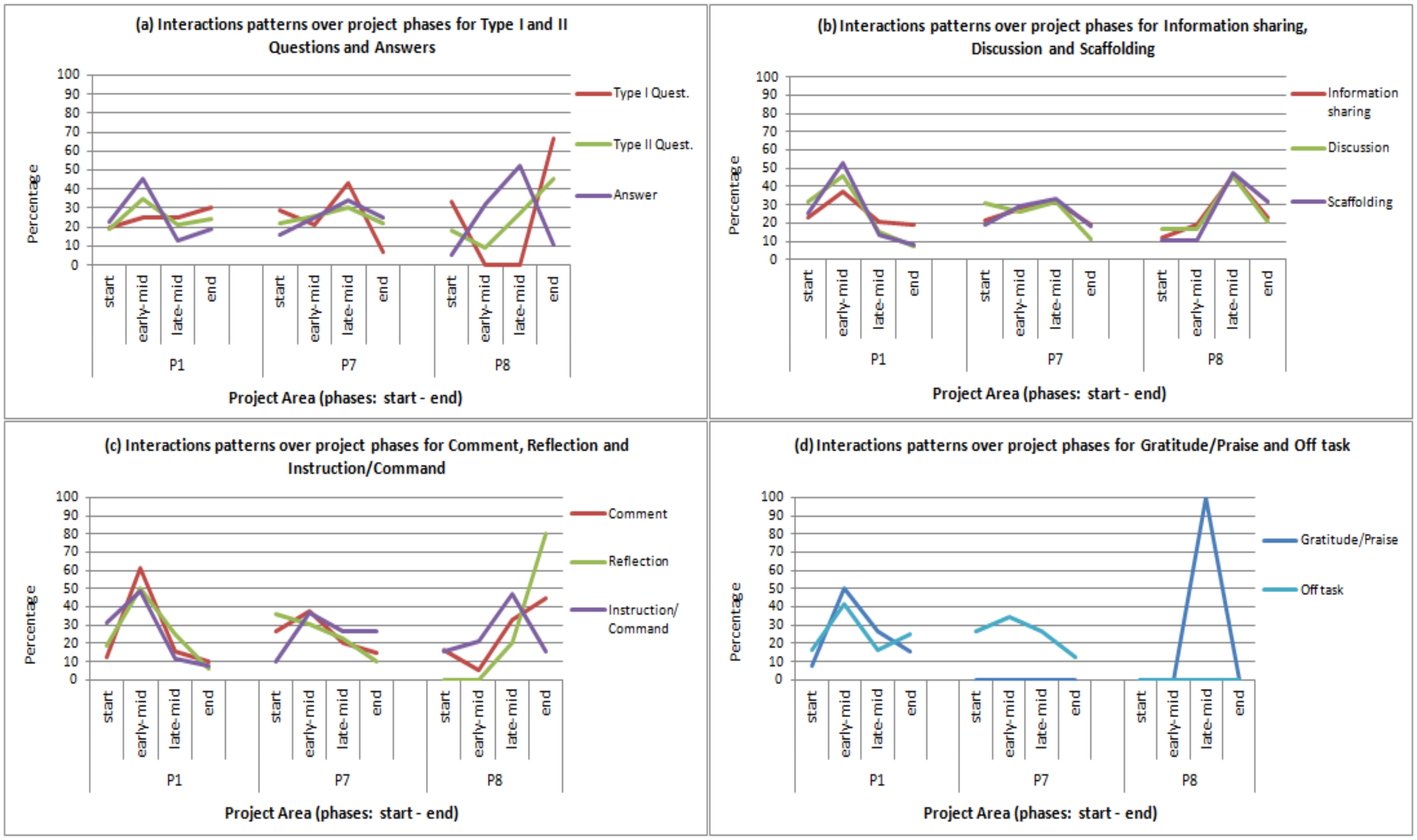

Figure 5. Detailed interactions of core developers over project phases for P1, P7 and P8

*Discussion of RQ2. What knowledge sharing behaviors do core developers exhibit over the course of their project?* **While core developers exhibited higher levels of some specific knowledge sharing behaviors at project start and completion, overall, these members shared more knowledge in the middle phases of their project.** Our

contextual analysis results show that the six Jazz core developers studied (from three project teams) were integrally involved in their project's knowledge sharing process [10], and that these practitioners were most dominant during the middle phases of their project. Additionally, these practitioners generally asked more questions at project start and project completion (although this pattern was not consistent for all three teams). A higher level of questioning at project initiation is understandable given the need to clarify and delineate system requirements and team understandings at this time, but this pattern is not necessarily expected at project completion. The elevated level of questioning at project completion coincides with core developers' reduced levels of project communication. The questions at this time may be aimed at the need for context awareness, given this reduced overall involvement. Our findings for Answers were the reverse; we observe that the six core developers in our sample provided most answers to their teams during the middle phases, when they were most active. These results suggest that although these practitioners were focused on their own individual responsibilities they also kept the teams' agendas in mind [80]. This assessment is supported by our other knowledge dimension analyses – this pattern of involvement is maintained for the six core developers' contributions of Information, ideas (Discussion) and suggestions (Scaffolding) to their teams, which were most frequently provided during the middle phases of the project. Information sharing was by far core developers' most dominant form of utterance, a pattern we linked to their task organisation and contextual awareness team roles (refer to [10] for detailed analyses of these patterns). Perhaps technical needs drove these core developers' knowledge sharing behaviors [1]; such an understanding was established by an early study considering human factors during software development [91]. Thus, communication hubs may not necessarily be formally denoted as communication and coordination specialists [1]. Rather (and accepting that our contextual analysis only considered the artifacts of six Jazz core developers), our findings provide a degree of evidence that these individuals communicate *because of* their teams' needs. While the presence of such individuals in software teams is no doubt beneficial, project managers should also be vigilant about the team's possible over-reliance on these members, which may negatively affect the quality of the knowledge core developers are able to provide. We extend our snapshot analyses [10] next to examine the way core developers contributed task changes over the course of their project.

## 4.3. Task Performance (RQ3)

In our prior investigation [10] we uncovered that core developers were indeed very active and made the most changes to their teams' tasks (noted previously). Here we report a more detailed analysis of how these changes were enacted in the individual project teams in order to answer RQ3. We then assess these members' actual contributions to task changes over time to see if these were correlated with their attitudes and knowledge sharing behaviors (in subsections 4.4 and 4.5, respectively).

Table 6 shows that core developers tended to make the most changes to their teams' tasks during the middle stages of their project. These individuals made as many as 33.3% of their changes in the early-mid phase for P2, and 47.0% of changes in the late-mid phase for P8 (see Table 6 for overall means which also maintain this pattern). The higher levels of task changes in these phases coincide with the higher numbers of messages communicated by these individuals, as noted in Table 4. Note also that we previously observed a small, positive correlation between the number of messages communicated and the number of task changes made by core developers [10].

We examine the performance of these individual core developers to see how they contributed in their given project teams over time in Figure 6. We observe a similar pattern to that evident in the overall results presented in Table 6. Apart from contributors 11643 and 12889 (these practitioners made 31.3% and 36.0% of their task changes in the end phase), all 13 other core developers exhibited reduced performance in task changes in the last phase of development. Figure 6 also illustrates that practitioners 13722 and 4674 made their highest contribution in the start phase (66.7% and 65.5% respectively), while the others were most active in the early-mid and late-mid project phases. We discuss these finding and answer RQ3 below.

Table 6. Percentage of overall task changes made by core developers over their project

| Team ID | Percentage of task changes | | | |
|---|---|---|---|---|
| | start | early-mid | late-mid | end |
| P1 | 19.4 | 26.2 | 25.2 | 29.1 |
| P2 | 19.0 | 33.3 | 28.2 | 19.5 |
| P3 | 35.3 | 14.7 | 32.4 | 17.6 |
| P4 | 20.8 | 26.9 | 25.6 | 26.6 |
| P5 | 28.9 | 19.1 | 37.5 | 14.5 |
| P6 | 27.2 | 26.5 | 25.7 | 20.5 |
| P7 | 22.1 | 32.6 | 23.2 | 22.1 |
| P8 | 11.6 | 20.5 | 47.0 | 20.9 |
| P9 | 20.8 | 26.0 | 32.5 | 20.8 |
| P10 | 19.5 | 27.4 | 30.3 | 22.8 |
| **mean** | **22.5** | **25.3** | **30.8** | **21.4** |

*Discussion of RQ3. How do core developers contribute to task performance over their project?* **Overall, Jazz core developers' involvement in task changes increased steadily over the first three project phases, and decreased towards project completion.** We previously uncovered that the core developers in our study initiated more than 41% of their teams' software tasks, they made more than 69% of the changes to these tasks, and resolved nearly 75% of all software tasks undertaken by their teams [10]. These findings are revealing when considering that core developers comprised just over 10% of their teams' 146 practitioners (see Section 3). In this study we consider core developers' involvement in task performance over the course of their project. As noted above, Jazz core developers were most active in the middle phases of their project (refer to Table 6 and Figure 6). In particular, these practitioners made the highest number of task changes in the late-mid stage of their project, with task changes tending to be stable, and much lower, during their project's start and end phases. Overall, core developers contributed the least of their task changes towards project completion. While this evidence was previously linked to the increasing level of task difficulty encountered as software projects progress [1], the trend of our results over the first three project phases does not support this position. It may instead be contended, given the associated reduction in communication, that core developers were involved in other work outside of these areas towards project completion. Future research employing complementary interview-related techniques would help us to understand the reasons for such a pattern. However, in the current work we are curious to understand how core developers' social motivation [54] impacted their task performance. Accordingly, we examine whether core developers' attitudes were linked to their task performance, as presented next.

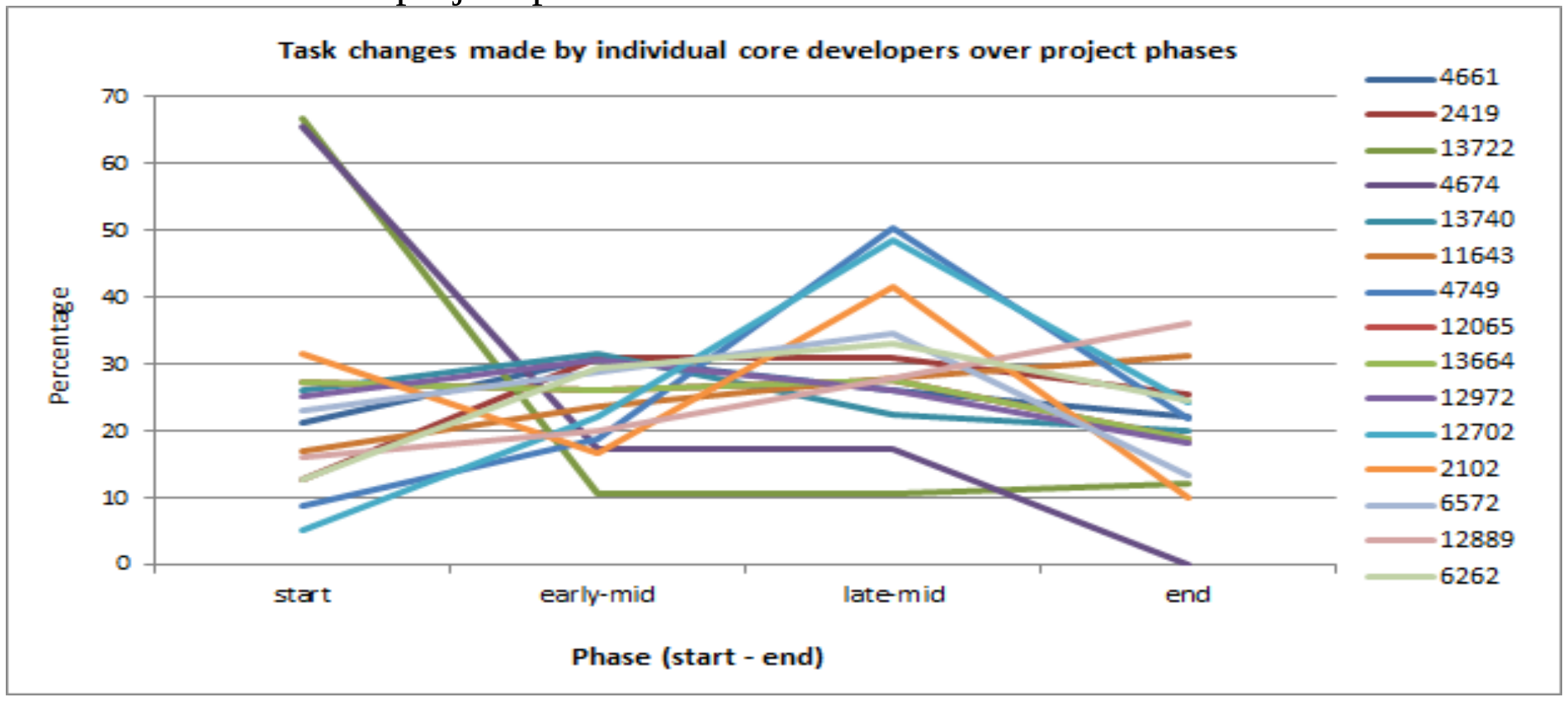

Figure 6. Percentage of task changes made by individual core developers over the course of their project

## 4.4. Attitudes and Task Performance (RQ4)

We conducted Pearson product-moment correlation analyses to address RQ4. In Section 4.1 we noted that core developers used more collective language (e.g., "we", "us", "our") towards project completion, but used a similarly higher proportion of individualistic language (e.g., "I", "me", "my") during the last project phase. While we did not find any relationship between individualistic language and the number of task changes, we found evidence of a small negative correlation between core developers' use of collective language and the number of changes they made, although this result was not statistically significant ($r = -0.211$, $n = 40$, $p > 0.05$) (refer to Table 7 for precise values). Similarly, although core developers were most active during the middle stages of their project, they also made the greatest use of reliance and delegation language (e.g., "you", "your", "you're") during this time. This pattern was also seen for insightful (e.g., "think", "consider", "determined") and discrepancy language (e.g., "should", "prefer", "needed") in Section 4.1. While there was no evidence of a relationship between the use of reliance language and the number of task changes, Table 7 shows that we found a small positive relationship between insightful language use by core developers and the number of changes they made ($r = 0.129$, $n = 40$, $p > 0.05$). In contrast, when core developers communicated with higher levels of discrepancy language they made fewer task changes ($r = -0.128$, $n = 40$, $p > 0.05$). Additionally, we observed the highest use of negative emotion language (e.g., "dislike", "suck", "stupid") towards project completion, and Table 6 shows that core developers were least active during this period. However, we did not observe any association between the use of this type of language dimension and the number of task changes made by core developers. While these relationships were not statistically significant, this could be a function of simply having too few data points – thus we have chosen to report them here as they warrant further investigation.

Table 7. Pearson product-moment correlation results for core developers attitudes and task performance

| Linguistic Category (Abbrev.) | Correlation Coefficient (r(40)) | Sig. (2-tailed) (*p*-value) |
|---|---|---|
| Pronouns (we) | -0.211 | 0.191 |
| Cognitive language (insight) | 0.129 | 0.428 |
| Cognitive language (discrep) | -0.128 | 0.431 |

*Discussion of RQ4. Are core developers' contributions to task performance linked to their attitudes?* **Our results do not provide a conclusive link between core developers' attitudes and their contributions to software tasks, although we observed some associations between task changes and the expression of some specific behaviors**. For instance, although we found a small negative correlation between our sample of core developers' use of collective language and the number of changes they made (refer to Table 7), this result was not significant. While it would be rational for core developers to express more collective attitudes [82] when relying on others for help (and this was observed in our results), this evidence was not strong. Opposite patterns of results were observed for the cognitive attitudes [73] in Table 7. When core developers communicated higher amounts of insightful language, they made more task changes, and when these individuals communicated with higher levels of discrepancy language they made fewer task changes. Insightful language comprises words such as "think", "consider", "determine", and "idea" [73]. It is expected that there would be elevated use of such words when assistance is being provided (e.g., "you should consider using session variables instead of cookies to maintain state across the web pages" or "I think the bug you are now noticing was observed after last night's build"), as was seen for core developers in the early-mid and late-mid phases of their project. Thus, this evidence provides preliminary support for our results that core developers were involved in higher levels of knowledge sharing when they made more task changes. Discrepancy language comprises words such as "should", "prefer", "needed", and "regardless" [73]. Such words are likely to be used to offer suggestions (e.g., "the patch I created for bug B should also work for bug C") or to show preference for a specific option (e.g., "I prefer option E over option F"). Thus, the finding that when core developers used higher amounts of discrepancy words they made fewer task changes is understandable. However, overall, our results in Table 7 are not definitive; this question would benefit from additional research over much larger samples of artifacts. We examine the way knowledge sharing was related to core developers' task performance next.

## 4.5. Knowledge Sharing and Task Performance (RQ5)

In order to answer RQ5 we examine the relationship between our directed CA results (in Section 4.2) and those obtained from our task change analysis (Section 4.3). We conducted Pearson product-moment correlation tests to determine the relationships between the knowledge sharing behaviors of core developers and the task changes they made; these results are presented in Table 8. The data show no violation of normality, linearity or homoscedasticity. Table 8 shows a medium, negative correlation between the number of Type I Questions asked (refer to Table 3 for example) and the number of task changes made, although this was not statistically significant ($r$ = -0.390, n = 12, $p$ > 0.05). In contrast, when core developers initiated more Type II Questions in their dialogues they made more task changes. Table 8 reveals that our correlation test to determine the relationship between the Answers provided by core developers and their involvement in task changes uncovered a strong, statistically significant positive correlation ($r$ = 0.691, n = 12, $p$ < 0.05). Similarly, our tests to determine the relationships between the contributions of Information, Discussion and Scaffolding by core developers and their involvement in task changes revealed strong, positive correlations for Information provision and Scaffolding, respectively ($r$ = 0.791, n = 12, $p$ < 0.01) ($r$ = 0.532, n = 12, $p$ > 0.05); however the result for Scaffolding was not significant. We found a medium, positive (but non-significant) correlation between core developers' contributions to Discussion and their involvement in task changes ($r$ = 0.466, n = 12, $p$ > 0.05) (refer to Table 8). We also found a strong, positive correlation between the volume of Instructions given by core developers (refer to Table 3 for example) and the number of changes they made, but this finding was also not statistically significant ($r$ = 0.539, n = 12, $p$ > 0.05). Our correlation test to determine the relationship between the number of Comments contributed by core developers and the task changes they made shows some evidence of a medium, positive correlation ($r$ = 0.308, n = 12, $p$ > 0.05), while core developers also expressed more Gratitude when they were involved in task changes (refer to Table 8). Taken together, associating our contextual analysis over time with our task change analysis confirms that core developers, although less active in the final project phase, were indeed above average performers, and these members were appropriately selected [10].

We triangulate the results of our investigation of the relationship between attitudes and knowledge sharing behaviors by conducting correlation tests to determine the relationship between core developers' use of cognitive linguistic dimensions (see "insight", "discrep", "tentat" and "certain" in Table 2) and their levels of contribution to Information, Discussion, Scaffolding and Comments (see Table 3). We found a strong, positive correlation between the incidence of insightful language use (e.g., "think", "consider", "determined") and core developers' contribution of Information (refer to Table 3 for example). This result was statistically significant ($r$ = 0.708, n = 12, $p$ < 0.05). We arrived at a similar but less conclusive finding for insightful language and Scaffolding ($r$ = 0.518, n = 12, $p$ > 0.05). Results for insightful language use and Discussion and Comments were less strong, but in these instances we also found positive correlations ($r$ = 0.470, n = 12, $p$ > 0.05) ($r$ = 0.345, n = 12, $p$ > 0.05). We did not find evidence of any relationships between the other cognitive dimensions (see Abbrev. "discrep", "tentat" and "certain" in Table 2) and core developers' contributions of Information, Discussion, Scaffolding and Comments. We consider these findings through the use of theory in the discussion that follows to answer RQ5.

Table 8. Pearson product-moment correlation results for core developers' knowledge sharing behaviors and task performance

| **Directed CA Category** | **Correlation Coefficient (r(12))** | **Sig. (2-tailed) (*p*-value)** |
|---|---|---|
| Type I Question | -0.390 | 0.210 |
| Type II Question | 0.215 | 0.501 |
| Answer | **0.691** | **0.013** |
| Information sharing | **0.791** | **0.002** |
| Discussion | 0.466 | 0.127 |
| Comment | 0.308 | 0.331 |
| Reflection | 0.073 | 0.821 |
| Scaffolding | 0.532 | 0.075 |
| Instruction/ Command | 0.539 | 0.710 |
| Gratitude/ Praise | **0.787** | **0.002** |
| Off task | 0.117 | 0.718 |

*Discussion of RQ5. Are core developers' contributions to task performance linked to their contributions of knowledge?* **Our findings show that Jazz core developers' performance in task changes is related to their contribution of some specific knowledge sharing behaviors.** Although this result was not statistically significant, we observed that when there were higher levels of questioning from the six core developers in our sample (for the contextual analysis) these practitioners were less active in task changes, tending to rely more on their wider project teams. Thus, strategies aimed at surrounding core developers with competent communicators may help these practitioners to quickly become familiar with task knowledge. This could in turn promote overall shared team understanding [95], and reduced incidence of coordination breakdowns [15]. This may be particularly useful for team synergies, as while it is not clear from our analysis why core developers demonstrated reduced presence at specific times of their project, results in Section 4.1 show that these individuals expressed more unhappiness towards project completion. Section 4.2 shows that, at this time, core developers also communicated increased levels of questions. Perhaps core developers were dissatisfied with the feedback (or lack thereof) they received [79]. In contrast, when core developers were actively involved in task changes they provided more answers to their wider project teams (refer to Table 8). A similar pattern of results was uncovered for Information sharing and Gratitude. Use of these forms of communication was related to core developers' active involvement in task changes. Scaffolding, although not recording a statistically significant finding, was also correlated with core developers' active involvement in task changes.

Overall our results lend support to previous psycholinguistic theories [73, 78] which established that there is evidence of individuals' skills, abilities and attitudes in their communications [80]. Thus, word use may indeed be contextual, and therefore, studying language patterns can reveal important details around team processes [96]. For example, we found that when practitioners used words such as "think", "consider", "determine", and "idea", they shared more ideas, offered more suggestions and were involved in more critical evaluations – and did more actual software development. These findings have implications for SE research and practice, considered in the next section (Section 5).

# 5. IMPLICATIONS FOR THEORY, PRACTICE AND TOOL DESIGN

In line with the previous findings of Bird et al. [4] our results confirm that core developers communicated the most in their project, and those that had frequent discourses were also integral to their teams' actual software development portfolios [1]. Overall, we observe that the sample of Jazz core developers studied demonstrated all competencies, including high levels of inter-personal, organizational and social skills [10]. Our results in this study suggest that core developers' competencies were enacted in different ways during different project phases, tending to align with the teams' needs and their involvement in task performance. While the findings revealed in this work are novel, we acknowledge that they are derived from the examination of artifacts of a small sample of core developers in a single case organization. Thus, we only provide conjectures for the patterns of results discussed in the previous section. We now consider the implications of our results for software engineering research (in subsection 5.1), software project governance (subsection 5.2) and collaboration and process tools (subsection 5.3).

## 5.1. Implications for Software Engineering Research

Previous studies have reported that a few individuals tend to dominate software projects [1-4], and it was also established that these individuals are crucial to their teams' dynamics [10]. This current work extends these studies by considering changes in these core developers' attitudes, knowledge sharing behaviors and task performance over the course of their project. Our result shows that these individuals expressed relatively stable attitudes over their project, and they were most integral to their teams' performance during the middle project phases. We also uncovered that core developers' performance in their teams' knowledge sharing networks was linked to their performance of task changes. Although previous work has examined core developers' contributions over time [29, 97], less emphasis has been placed on examining how core developers' organizational, inter-personal and intra-personal competencies sustain their project's health, and particularly taking an in-depth view of core developers' attitudes and knowledge sharing processes.

Additionally, the value of employing more contextual analysis techniques to understand team processes cannot be overstated. Studies in the SE discipline examining human processes using communication artifacts and repository data have tended to employ analytical and frequency-based approaches [4, 20]. Such approaches align with a technical focus on team processes [98]. Although certainly useful, it is generally understood (and there is growing recognition in SE) that these approaches would benefit from triangulation with more contextual techniques [28, 99]. This view is also supported by researchers working in the Information Systems (IS) discipline [100], where tested research approaches in the behavioral sciences, management and psychology domains have been recommended for use when studying human aspects of processes [101]. For instance, in terms of behaviors and attitudes, linguistic analysis tools (such as the LIWC tool that was used in this study) are said to capture individuals' behaviors by assessing the types of words they use [73, 102]. Thus, complementing frequency based SNA metrics [4, 5, 13, 30, 32, 43] with this form of analysis would provide further insights into software teams' human processes . In fact, although it may be argued that counts of words may not capture the *specific* contextual meaning of word usage [103, 104], when taken together the usage of such words can indicate an individual's temperament, their language composition preferences, their psychological traits [73] and their moods [36]. Other forms of thematic and intuitive analysis (such as the directed CA that was used in this study) may complement word-based analysis and help us to understand the reason for, and effects of, word use [105].

Furthermore, previous work has placed significant emphasis on the need for longitudinal studies to understand changes in teams' activities over time, as against the frequently used snapshot or cross-sectional analysis approaches [13, 32]. Our utilization of deeper analysis techniques in this longitudinal study indeed shows that studying software practitioners' behaviors, even at the word usage level, provides enhanced understanding of SE project teams [106]. This could be readily evaluated from the results and analysis in Section 4. While we did not study core developers' communication network properties over time (rather, we studied their contribution of messages in Section 4), perhaps specific patterns in these members' networks are linked to their word use. Core developers' communication networks are likely to be less dense and their closeness measures higher (noting that higher measures for network closeness indicate that members are less reachable) during the latter project phase, as per their reduction in communication as noted in Table 4. These patterns may be linked to core developers' higher use of individualistic language (e.g., "I", "me", "my"), as was noted in Table 5 when there were lower levels of communication. Similarly, higher levels of network closeness (so lower measures) may be linked to higher levels of productivity, as we noted for core developers' use of insightful language (e.g., "think", "consider", "determined"), contribution of knowledge sharing behaviors and task changes in Section 4.5 – core developers' use of insightful language was found to be related to their contribution of Information, and their contribution of Information was related to their involvement in task changes. We encourage those exploring the work of highly productive software practitioners, and SE team issues in general, to conduct such temporal analyses and to triangulate frequency-based approaches with contextual analysis techniques.

A further step in this work, had it been possible, would be to conduct interviews with these core developers to understand the details of their motivation and the reasons for their less pronounced presence towards project completion. Accordingly, we encourage future research to use such an approach to examine this issue. Specifically, we encourage research to address the following open questions: What ignites (or dampens) core developers' motivation? Why are core developers most negative and self-focused at project completion? What is responsible for core developers' reduced task performance and communication towards project completion? Research may also consider the questions addressed in this study in relation to the quality of core developers' deliverables (e.g., How do core developers' attitudes and knowledge sharing behaviors affect the quality of the features they deliver?).

## 5.2. Implications for Software Project Governance

The core developers studied here were found to be integral to their project teams. These practitioners exhibited higher levels of cognitive processes at project initiation (and in the middle phases), which may benefit weaker team members, and so it would be prudent for project managers to implement strategies aimed at encouraging the engagement of less active developers at project inception or in the earlier parts of the project. On the other hand, core developers were most negative and self-focused at project completion. This result coincides with evidence of reduced levels of participation in task changes and communication at this time, suggesting that these practitioners' dependence on others may have been responsible for their expression of less desirable behaviors. Thus, project managers may need to encourage communication and facilitate the availability of the wider team towards project completion (or whenever there is reduced visibility of core developers) in order to reduce tension and potential fallouts in the team. Strategies aimed at surrounding core developers with other competent communicators would likely also help core developers to quickly become familiar with current task-related knowledge.

While core developers led their project teams overall [10], these practitioners maintained particularly strong task and communication performance in the middle phases of their project. At this time, these members provided most of their answers, information, ideas and suggestions to their wider project teams. This in turn perhaps caused their teams to rely heavily on them for task knowledge; and this reliance likely impacted core developers' distinct presence. Given that core developers are the conduit for team knowledge, project managers should be vigilant in ensuring that they are supported by a wider team who are willing to ask questions and challenge the core developers' opinions and ideas. This is particularly necessary given core developers' heavy workload, which may have a negative impact on the quality of the knowledge they provide.

Finally, our findings indicate that these core developers communicated *because of* their actual development portfolio; the more task changes these practitioners made, the more messages they communicated. The extent of the exchanges in which core developers engage has implications for the availability of communication channels. This is especially critical for distributed project teams (such as those studied here), as most of the content communicated by core developers related to their work on software tasks. Project managers should be vigilant about the likely negative effect of temporal distance [39] on core developers' need to communicate extensively.

## 5.3. Implications for Collaboration and Process Tools

The findings of this study also have implications for the enhancement of collaboration and software process support tools. Our results show that core developers' communication increased with higher levels of task involvement and their attitudes and knowledge sharing behaviors varied over the phases of their project. The link between communication and task involvement suggests that tools may provide useful visualizations of measures of development and coordination carried out by practitioners during the software process. These project metrics may be compared against perceived or projected coordination measures in support of team management [107]. A project manager could use such tools in a similar way that project management and tracking tools are used for monitoring project performance. (Such tools would need to be sufficiently informed about the linkage between practitioners' involvement in task changes and communication.)

Process tools that capture practitioners' communications (such as IBM Rational Jazz) may also benefit from attitude and behavior visualizations. Perhaps such a feature may also be helpful for OSS teams, and teams globally distributed in general, where it would be prudent to manage team behaviors given team members' separation, and thus, the need to maintain cohesive and balanced team morale [108, 109]. In order to be reliable and accurate, such tools would need to adhere to sound data mining principles (particularly data pre-processing techniques) and tested natural language processing methods [66]. These tools could help project managers with team composition and task assignment [110]. For instance, in monitoring teams' attitudes over the course of their project a relatively high incidence of negative words (e.g., "afraid", "hate", "suck", "dislike") expressed among developers would be an indicator of frustration and dissatisfaction. Such an observation may kick start human resource interventions (e.g., deeper interviews which lead to increasing the complement of highly skilled developers, staff rotation or some form of team-building activity).

Finally, our results show that the majority of these core developers' communications were contributed towards actual problem solving. However, some messages were more urgent than others (e.g., the questions asked towards project completion). In order to manage the wealth of communication that flows within teams, particularly in distributed projects such as this, a message tagging feature could be included in Jazz or similar tools (as is done for tagging software tasks). In this way, critical communication could be labeled accordingly. During time-constrained work periods, comment tags should enable practitioners to identify and consider the most critical issues first. This would permit project teams to better manage their communication and feedback.

## 6. LIMITATIONS

*Measurement Validity*: The LIWC language constructs used to measure attitudes in this study have been utilized previously to investigate this subject and were assessed for validity and reliability [78]. However, although the LIWC dictionary was able to capture 66% of the overall words used by Jazz practitioners, the adequacy of these constructs in the specific context of software development warrants further investigation. To that end, we checked a small sample of the messages to see what might account for the remaining words being ignored by the LIWC tool and found that there were large amounts of cross referencing to other WIs in the messages, along with large amounts of highly specialized software related language (e.g., J2EE, LDAP, JACC, API, XML, TAME, JASS, Jazz, URI, REST, HTTP, Servlet, WIKI, UseCase, HTML, CVS, Dump, Config, SourceControl) evident in Jazz practitioners' exchanges. Their non-consideration here is therefore not a problem. Moreover, what was of interest, and was captured by the LIWC tool, was evidence of attitude, demeanor and behavior. Our linguistic results were also triangulated with positive and significant correlation results from our contextual analysis.

Communication was assessed based on messages sent around software tasks. These messages were extracted from Jazz, and so, may not represent all of the project teams' communications. Offsetting this concern is the fact that, as Jazz was developed as a globally distributed project, developers were required to use messages so that all other contributors (irrespective of their physical location) were aware of product and process decisions regarding each WI [32].

Our directed CA involving interpretation of textual data is subjective, and so questions may naturally arise regarding the validity and reliability of the outcomes of this analysis. In this work we employed multiple strategies to mitigate these issues. First, our protocol was adapted from those previously employed and tested in the study of interaction and knowledge sharing [84, 85], and so there is a strong theoretical basis for its use. Second, we piloted the protocol and extended our instrument by deriving additional codes directly from the Jazz data and we tested this extended protocol for accuracy, precision and objectivity, receiving an inter-rater measure indicative of excellent agreement [86]. However, we believe that additional research employing interview techniques is necessary to further validate the outcomes from these analyses.

Cultural differences and distance (geographical and temporal) may directly affect software development teams' performance [39], and these variables may also have an impact on team members' behaviors - which in turn may lead to performance issues [61]. However, as noted in Section 3.1, research examining the effects of cultural differences in global software teams has found few cultural gaps and behavioral differences among software practitioners from, and operating in, Western cultures, with the largest negative effects observed between Asian and Western cultures [39]. Given that the teams studied in this work all operated in Western cultures, this issue may have had little effect on the patterns of behaviors that were observed. Further, our focus in this work was on core developers, as against the team.

Finally, we use task changes (including measures for task created, task modified and task resolved) to determine core developers' task performance [1]. However, all software tasks are not equal [45]; some tasks may be more computational and complex than others (e.g., a user experience tasks may not demand the same cognitive and mental rigor as that of a computational or coding-intensive feature). Nonetheless, such complexities would likely 'even out' over the entire project. Additionally, when these measures are compared across the Jazz project teams studied, the results show that core developers were integrally involved in their teams' tasks, regardless of the task type [10]. Further, our in-depth analysis of core developers' messages triangulates these measures, and confirms that the members selected as core developers were indeed top performers.

*Internal and External Validity*: Although we achieved data saturation after analyzing the third project case, the tasks, history logs and messages from the ten project teams (out of 94) may not necessarily represent all the project teams' processes in the repository. Additionally, we studied teams from a single organization employing particular development practices. Work processes and work culture at IBM are likely to be specific to that organization and may not be representative of organization dynamics elsewhere, and particularly for environments that employ conventional waterfall processes [111]. Such environments may employ more rigid project management practices, with clear hierarchical structures, development boundaries and defined roles [112-114]. Accordingly, this form of project administration is likely to restrict core developers' performance, and particularly, their ability to operate fluidly across roles, or for them to even demonstrate such a distinct presence. That said, Costa et al. [115] confirmed that practitioners of the Jazz project exhibited similar coordination needs to practitioners of four projects operating in two distinct companies. Thus, we believe that our results may be applicable to similar large-scale distributed projects.

## 7. CONCLUSION

This study contributes to our understanding of core developers' communication processes, and how these relate to their task performance. Our findings indicate that core developers' attitudes and involvement in knowledge sharing were linked to the demands of their wider project teams. However, these practitioners also brought high levels of skills and cognitive characteristics to their project teams. These individuals started their project providing high levels of ideas, suggestions, information, comments, instructions and answers to their teams, and became the center of their project's knowledge activities as the project progressed. These patterns were related to core developers' actual involvement in task changes – the more changes core developers performed the more knowledge they provided. When these practitioners were least involved in communication and task changes, they exhibited some negative team attitudes, suggesting that they could have been dissatisfied with the feedback (or lack thereof) from the wider team. These findings have implications for future research, software project governance and collaboration and process tool enhancements.

## ACKNOWLEDGMENTS

We thank IBM for granting us access to the Jazz repository, and the coders for their help in the data analysis. Thanks to the reviewers for their detailed and insightful comments on the early versions of this work, and Daniela Damian for her direction. S. Licorish carried out the research underpinning this paper while supported by an AUT University Vice-Chancellor's Doctoral Scholarship Award and a Post Submission Doctoral Scheme Award.